\documentclass[%
 reprint,
 amsmath,amssymb,
 aps,nofootinbib,   
]{revtex4-1}
\usepackage{bm}
\PassOptionsToPackage{linktocpage}{hyperref}
\usepackage[hyperindex,breaklinks]{hyperref}
\usepackage{enumitem}
\usepackage{slashed}

\renewcommand{\theta}{\vartheta}

\renewcommand{\vec}[1]{\ensuremath{\boldsymbol{#1}}}

\newcommand{\bra}[1]{\ensuremath{\left< #1\,\right|}}
\newcommand{\ket}[1]{\ensuremath{\left|\, #1\right>}}

\usepackage{array}
\usepackage{mathtools}

\usepackage{etoolbox}

\begin{document} 

\title{Swift Memory Burden in Merging Black Holes:  \\
 {\it how information load affects black hole's classical dynamics}}

\author{Gia Dvali } 
\affiliation{%
Arnold Sommerfeld Center, Ludwig-Maximilians-University,  Munich, Germany, 
}%
 \affiliation{%
Max-Planck-Institute for Physics, Munich, Germany
}%

\date{\today}

\begin{abstract} 

 In this paper we argue that the information load carried by a black hole affects its classical perturbations.  
  We refer to this phenomenon as the ``swift memory burden effect" and show that  it is universal for objects of high efficiency of information storage. The effect is expected 
 to have observable manifestations, for example, in mergers of astrophysical black holes in Einstein gravity.  
  The black holes with different information loads, although degenerate in the ground state, respond very differently to perturbations.  
  The strength of the imprint is controlled by the 
 memory burden parameter  which measures the fraction of the black hole's memory space occupied by the information load.  This represents a new macroscopic quantum characteristics of a black hole.
    We develop a calculable  theoretical framework 
  and derive some master formulas which we then test on explicit models  of black holes as well as on 
solitons  of high capacity of information storage.  
 We show that the effect must be significant for the spectroscopy of both astrophysical and primordial black holes and can be potentially probed in gravitational wave experiments.    
   We also provide a proposal for the test of the memory burden phenomenon  in a table-top laboratory setting with cold bosons. 

  \end{abstract}

\maketitle

\section{Introduction} 

 In this paper we shall discuss a novel manifestation 
 of the memory burden effect \cite{Dvali:2018xpy, Dvali:2018ytn, Dvali:2020wft}, to which we shall 
 refer as the ``swift memory burden" phenomenon. 
 
  The effect is generic for the system 
 of  high efficiency of information storage, such as black holes, and affects their classical dynamics.  In particular, it must be operative in the mergers of ordinary astrophysical black holes within the framework of Einstein gravity.  
 Our key message is:\\
 
  {\it  The information load carried by a black hole affects its classical dynamics when the black hole is perturbed.} \\

  It is a common knowledge that  black holes are the most compact information storages.  Despite this, ordinarily, the quantum information load carried by a macroscopic black hole is not accounted in its dynamics. 
 Instead, it is treated as sort of an exotic feeble entity  
 playing no role in physics of large black holes. 
 
 This view is completely opposed by the memory burden 
 effect  \cite{Dvali:2018xpy, Dvali:2018ytn, Dvali:2020wft}.  
  The essence of the generic memory burden
  phenomenon  is 
  that the information (or memory) load carried by a system 
 affects its time-evolution.  Specifically,  the back-reaction from the memory load  resists to a departure of the system 
  from the state of high capacity of information storage.  The phenomenon is 
 generic and is exhibited by a large universality class 
 of objects of high efficiency of information storage.
 
  Such systems universally employ the 
 mechanism of the ``assisted gaplessness"  
 \cite{Dvali:2017nis, Dvali:2017ktv, Dvali:2018vvx, Dvali:2018xoc, Dvali:2018tqi}, \cite{Dvali:2018xpy, Dvali:2018ytn, Dvali:2020wft}.  The meaning is that the system is
 creating a local environment in which the information-carrying 
degrees of freedom, the so-called ``memory modes", become gapless. As a result, they can store large loads of information at a very low energy cost.   
   Thus, the state of the assisted gaplessness represents  the state of a high efficiency of information storage and the  two terms can be used interchangeably. 
   The information is stored in various patterns of excitations of the memory modes, ``memory patters".  These states span the ``memory space".

 Now, when the memory modes are loaded,  this creates an energy barrier resisting the system to move 
 away from the gaplessness point. This is the essence 
 of the memory burden effect  \cite{Dvali:2018xpy, Dvali:2018ytn, Dvali:2020wft}.   
 
   Although the black holes are the most prominent 
 representatives of the above universality class, the features are shared 
 by  all quantum field theoretic  (QFT) objects of the high  microstate degeneracy, such as the saturated solitons (the so-called ``saturons"  \cite{Dvali:2020wqi}) \cite{Dvali:2019jjw, Dvali:2019ulr, Dvali:2020wqi, Dvali:2021ooc, Dvali:2021jto, Dvali:2021rlf, Dvali:2021tez, Dvali:2021ofp, Dvali:2023qlk, Dvali:2024hsb, Contri:2025eod}.  Correspondingly, such solitons  are subjected to the memory burden effect 
 \cite{Dvali:2021tez, Dvali:2024hsb}.

  Already in the first paper on this subject   \cite{Dvali:2018xpy},  it was shown that the 
  system resists to arbitrary departures
  from the state of efficient  information-storage.
This applies equally to a quantum decay as well as to  a coherent classical evolution.     
  In both cases the dynamics is strongly affected by 
  the information load carried by the system. 
   
  However, until now, the main implications of the memory burden in  black holes were discussed in the context of 
 the  back-reaction to the Hawking evaporation.  
  That is, an evaporating black hole gradually enters the memory burden phase beyond which the 
  semi-classical regime is no longer applicable. 
   Studies show that beyond this point the decay 
   rate slows down and the life-time gets prolonged.   
    This can have variety of important implications, in particular, for primordial black holes (PBH).    

 The characteristic feature of the memory burden 
 regime reached via a quantum decay is that it sets in gradually over a macroscopic time. 
  Correspondingly, such memory burden can have observational implication only for sufficiently light black holes that had enough time to enter the burdened phase.  
    
  In the present paper we point out that  the memory burden can generically be activated on much shorted time scale due to an external classical disturbance experienced by the black hole.  Such a disturbance can occur, for example,  during a merger with another massive objects.   
    Under such perturbations, the swift memory burden  must be experienced by a black hole of arbitrarily high mass.  
    This fact sharply increases the observational value  
    of the phenomenon.

     Black holes of the same mass have equal 
    information storage capacities but their actual information
    loads are in general different.   In the black hole's  ground state the information load is not classically-observable. 
  However,  it gets activated in form of the memory burden effect  as soon as the black hole is perturbed. 
     Correspondingly, the black holes 
     with different information loads are expected to exhibit very distinct classical evolutions.  
  
    A situation of this sort naturally takes place during the black hole mergers.   
   We shall argue that in such a case the memory burden is activated swiftly and affects the classical dynamics of the merger.   
   The strength of the swift memory burden depends 
  on the information load carried by a black hole. 
  This load is inherited  from the collapsing source that produced a given black hole. 
  
   We formulate a calculable framework that allows us 
   to  account for the essential dynamics and make 
   some model-independent predictions.   
    In order to parameterize the strength of the 
    effect, we introduce a black hole memory burden parameter $\mu$. This quantity controls the fraction  
of the black hole's information capacity occupied by its actual information load. We are thus lead to the following 
statement.\\

 { \it A classical black hole,  beyond its ordinary
 features, such as the mass,  the charge and the angular momentum, carries an additional macroscopic characteristics in form 
 of the memory burden parameter, $\mu$, 
 which measures the fraction of the black hole's  memory space used by its information load.  This quantity  
 has (almost) no effect in the black hole's  ground state 
  but gets  activated for a perturbed black hole,  influencing 
  its dynamics.}   \\

The strength of the memory burden effect is stronger 
for smaller $\mu$.   
   We estimate that for astrophysical black holes 
  obtained by an ordinary collapsing matter, 
 $\mu \ll 1$. This implies that in mergers of such 
 black holes, the  classical dynamics must be affected significantly. 
  These departures from ordinary evolution  have sufficient strengths to  be potentially observed via  their imprints in gravitational waves.

    The  discussion is  structured in the following 
    steps.  Before entering the technical details, we give a general physical argument explaining, at an intuitive level, 
 why the classical dynamics must be sensitive to the information load.  After this intuitive discussion, we move to  substantiating it.     
  
   First, we formulate a calculable framework in which  we carefully define the universality class of objects 
    of the efficient information storage and 
    highlight some important features.     
     We then introduce the 
    memory burden effect and its swift regime. 
     We derive some key formulas which we then apply 
     to black holes.  We come up with  some general predictions for black hole spectroscopy. 
              
      Next we discuss the swift memory burden effect in solitons. Finally we outline a proposal  for testing the memory burden effect  in table-top labs. We use the illustrative example of a  systems with attractive cold bosons.    
  
\section{A physical argument} 

  Before moving to a more technical part of the paper,  we would like to offer an intuitive argument  indicating why the behaviour of a perturbed  system must be affected by its information load. 
 The argument  is applicable to a generic 
 system of high efficiency of information storage. 
 However,  for concreteness and familiarity,  
 we shall formulate it for a black hole.   
 Every concept entering this discussion will be
  defined and quantified in the rest of the paper. The purpose of outlining the argument here is the  
 preparation of an intuitive base.   
  
   The argument was previously 
   given in  \cite{Dvali:2018xpy}, and is based on a
   comparison of  the energy costs of one and the same information
 pattern inside and outside of a black hole.  
   
 Let us consider a black hole  of mass $M$. 
  The radius of it is $R\sim M/M_P^2$, where $M_P$ is the Planck mass.  This black hole can carry an exponentially 
  large variety of information patterns, i.e., the memory patterns. Their diversity is 
   $\sim {\rm e}^S$, where $S \sim  M^2/M_P^2$ is the 
   Bekentein-Hawking microstate entropy. 
   These patterns can have different information loads, i.e.,  
 can use up  the different portions of the black hole's memory space.  This is determined by the number of
 excited memory modes forming the pattern. 
   
 For definiteness, let us assume that  a given black hole carries  a memory pattern  of 
 a near-maximal information load.  The number of
 memory modes involved in such a pattern is 
 $\sim S$.  
 
  In the absence of the black hole,
 the energy cost  $E_p$ of this information pattern,  
 encoded in modes of a free quantum field localized in  a box of radius 
 $R$,  would  exceed the mass of the black hole 
 by a factor $M/M_P$, 
  \begin{equation}  \label{AAA}
   E_p \, = \, M \frac{M}{M_P} \,, 
  \end{equation}    
   For example, for a solar mass black hole, this factor 
   is $10^{33}$.   In order to appreciate the magnitude 
   of the effect, notice that the energy of such information pattern would exceed the mass of the entire visible universe by many orders of magnitude. 
 
   In contrast, the black hole manages to store 
   this information at essentially zero energy cost. This is clear from the fact that the black holes with different information loads have the same mass.  
   
   Thus, the black hole possesses  a mechanism that minimizes the energy cost of the pattern. We shall quantify the microscopic  origin of this mechanism later. However, 
 for the present argument this is unimportant. 
 It suffices to know that  in the black hole the energy of the information pattern is minimized from (\ref{AAA}) to  
 (almost) zero.  That is, the  black hole creates a local environment 
 that nullifies the energy costs of arbitrary patterns. 
 
  Already at this point it is intuitively clear that 
  such an extraordinary tradeoff  cannot stay indifferent 
  to perturbations of  the system. 

  It is useful to picture the above tremendous energy-difference as a potential energy landscape with the bottom at zero.  This bottom is a flat ``valley" which accommodates 
  all possible patterns.   When the black hole is in its ground state,  the information pattern is at the bottom of the potential,  regardless of  the size 
 of its information content.  In other words, in black hole  ground state,  all information patterns 
 are degenerate in energy.  
 
 Due to this, unperturbed 
black holes are degenerate in mass, irrespective of 
the information loads they carry. This degeneracy is also the origin of the black hole's micro-state  entropy $S$. 
 For an unperturbed classical black hole it is impossible to tell what is the information  (memory) load that it carries.  

 However, the steepness and the hight of the walls of the valley depend on the information load of the pattern. 
 For patterns containing more information, the walls are steeper and higher (see cartoon in Fig. 1., of  \cite{Dvali:2018xpy}). 
 
  When a black hole is perturbed,  the memory load is pushed away from the minimum. This requires climbing up the slope of  the potential.  This creates  the back-reaction force 
  which is the  essence of the memory burden effect  \cite{Dvali:2018xpy, Dvali:2018ytn, Dvali:2020wft}. 
  The time-scale  of the burden is set by the time-scale of perturbation.   
  
  For a gradual decay, such as the Hawking 
  evaporation, the burden sets in over a macroscopic time.  
  In contrast, for classical perturbations the effect 
is expected to be swift  and to influence the classical dynamics of the system. 
  
   The question is whether, for certain perturbations, the memory burden can be avoided. For perturbations directed towards the decrease of the information storage capacity, such as  the Hawking evaporation,  the memory burden
is unavoidable.

     As for a generic classical perturbation, in order for the memory burden to be avoided, the entire process must proceed without  influencing  the 
  memory modes.   In other words, coincidentally, the evolution 
  must  take place exactly on a null memory burden surface.
    As we shall discuss, this appears highly unlikely.
 
 In the rest of the paper we  shall step by step deconstruct and quantify the phenomenon.

  \section{Setting the framework} 

 For a long time black holes have been considered 
 as one of the most mysterious objects of nature, 
 unique in their own category. 
 This is mainly due to their properties of information processing. The following features are well-established:

  \begin{itemize}
  \item  For a given size, the black holes have maximal  information storage capacity, 
   expressed in Bekenstein-Hawking  microstate entropy 
   \cite{Bekenstein:1973ur, Hawking:1974sw}. 
   For a black hole of radius $R$ in $d$ space-time dimensions this entropy scales as the surface area in Planck units, 
   \begin{equation} \label{BekH}
   S \,  \sim \, (R M_P)^{d-2} \,.  
\end{equation}
  
  \item  Semi-classically, black holes possess information horizons;

 \item   Within the validity of semi-classical treatment, black holes   decay  via Hawking evaporation \cite{Hawking:1974sw}, 
   emitting particles in an approximately thermal  spectrum, with the temperature set  by the inverse radius, 
       \begin{equation} \label{Hawking}
    T  \sim \frac{1}{R}  \,.  
\end{equation}

 \item  Despite being  enormous information reservoirs, 
 semi-classically, black holes emit pure energy that  carries  essentially no information content.

\end{itemize}
    
 Despite being well-established within the semi-classical treatment, the above features sometimes are
considered to be ``mysterious" due to a lack of commonly 
accepted microscopic explanation behind them.

    For addressing this question, the first task in the present paper  will be to establish the domain of validity of the semi-classical regime and to outline the strategy for 
     identifying the important effects that have not been captured within this domain.    \\  
   
   In the standard treatment of extrapolation of 
   Hawking's semi-classical  result to a finite mass black hole,  the black hole evaporation is assumed to be  self-similar.
   
   That is,  after emitting the energy  $\Delta M = M-M'$, the black hole of initial mass $M$ becomes a lighter and smaller black hole with the identical relations between  its mass $M'$,  the radius $R'$, the temperature $T'$ and the entropy $S'$:
    \begin{equation} 
       R' = \frac{M'}{M_P^2}\,,~~~\, T' = \frac{1}{R'} \,,
       ~~~  S'= \frac{(M')^2}{M_P^2} \,,
     \end{equation} 
   (the irrelevant numerical factors shall be ignored). 
 However,  as pointed out in \cite{Dvali:2015aja}, this assumption is  self-contradictory because the above relations would also imply,  
 \begin{equation}  \label{Gbound}
     \frac{\dot{T}}{T^2} = \frac{1}{S} \,,
 \end{equation} 
  where dot stands for the time-derivative. 
  The parameter $\frac{\dot{T}}{T^2}$ represents the 
measure of validity of thermal approximation.  
  Correspondingly, the above equation puts a lower bound on the precision of validity of Hawking's semi-classical regime.  That is, with each emission, the black hole experiences a quantum back-reaction of order $1/S$.  
 
  At first glance,  for a macroscopic  black hole, this 
  correction may look insignificant, since $S \gg 1$. 
 However, one must remember that the (naive) half-decay
 time  of a black hole is proportional to its entropy, 
  \begin{equation} \label{Thalf}
     t_{\rm half}  = S R\,.
 \end{equation} 
Correspondingly, first, even if nothing happens before, 
the corrections accumulated 
 over the time $t_{\rm half}$ can be significant. 
  Secondly, the formula (\ref{Gbound}) only 
  puts the lower bound on the validity of the semi-classical treatment. In reality, the corrections can grow much faster. 
  As we shall see, this is likely the case.   
   
    Another  killer argument   \cite{Dvali:2015aja, Dvali:2018xpy, Alexandre:2024nuo} against self-similarity of the black hole decay is coming from the entropy.  If evaporation is self-similar, then after reducing the mass say by a half, the entropy  would decrease by a factor of four,  
      \begin{equation} \label{Shalf}
     M \rightarrow  M'=\frac{M}{2}\,,~~~ S \rightarrow S' = \frac{S}{4}  \,.
     \end{equation} 
 At the same time, within the validity of the semi-classical regime the information is not coming out. This gives an obvious contradiction, since 
  there is no way to accommodate the maximal information load carried  by the initial black hole within a black hole of the reduced entropy.   
     
  The above arguments are very powerful as they 
  are derived solely under the assumption of semi-classicality  and show the lack of self-consistency.  
  
   We thus see that the black hole evaporation process cannot be self-similar and some departures must happen from it.  In order to understand what actually happens, the quantum  back-reaction must be taken into account.
   This requires knowledge of the microscopic mechanism. There are the following two ways of achieving this.  \\
   
   \subsubsection{Microscopic theory} 
   
     The first approach is to develop a microscopic theory that 
 gives a calculable framework for accounting the back reaction.  Such a proposal was put forward in \cite{Dvali:2011aa, Dvali:2012rt, Dvali:2012en, Dvali:2012wq, Dvali:2013eja, Dvali:2013vxa, Dvali:2014ila, Dvali:2015ywa, Dvali:2015wca, Dvali:2016zqx, Averin:2016hhm}  and goes under the name of  ``black hole's quantum  $N$-portrait". In this framework black hole is described  as a coherent state (or a condensate) of gravitons at  the point of quantum criticality \cite{Dvali:2012en}.
   In this picture, the Hawking 
 evaporation is described as the quantum decay of the graviton coherent state due to their re-scattering. Correspondingly,  the back-reaction at the initial stages 
 has been shown to be indeed $\sim 1/S$. 
 
 In the present paper we shall not use the above microscopic 
 portrait as our starting point. Nevertheless, it shall be useful in two ways.  First, the general phenomenon of memory burden  \cite{Dvali:2018xpy, Dvali:2020wft}, when applied to black holes,  naturally supports their composite picture. Secondly, the intuition obtained from $N$-portrait, shall be useful in identifying how the memory  burden effect operates in black holes at a microscopic level. 
    
 \subsubsection{Benefiting from universality of the phenomenon} 
      
     In this paper, we primarily focus on the program which can be described as: \\
     
   {\it gaining insights into the microscopic 
     picture from the universal nature of the phenomenon}.  \\
     
  This approach was developed in series of papers, 
  from two different angles:  \\
  
 {\it 1)} Defining the generic systems of efficient information storage and identifying the universal underlying mechanisms
 \cite{Dvali:2017nis, Dvali:2017ktv, Dvali:2018vvx, Dvali:2018xoc, Dvali:2018xpy, Dvali:2018tqi}.  Exploring new accompanying phenomena such as the ``memory burden" effect \cite{ Dvali:2018xpy, Dvali:2020wft};  
   
  and 
 
  {\it 2)}  Deriving QFT bounds on information capacity and 
  studying the universal features of systems that saturate this capacity 
   \cite{Dvali:2019jjw, Dvali:2019ulr, Dvali:2020wqi,
Dvali:2021ooc, Dvali:2021jto, Dvali:2021rlf, Dvali:2021tez, Dvali:2021ofp, Dvali:2023qlk, Dvali:2024hsb}.    \\

    A scientific method for understanding a mysterious phenomenon exhibited by a particular system 
    is to ask whether the same phenomenon repeats itself 
    in other systems.   If this is the case, then there is 
    a strong indication that the underlying microscopic 
    mechanism is universal. 
    
    The universality endows physicists  with a  great power:
  on one hand,  it allows to study the phenomenon in the systems  that are more calculable, and, on the other hand,  it makes possible to predict new effects. 
    
     The example of such new phenomenon is provided  by the memory burden effect \cite{ Dvali:2018xpy, Dvali:2020wft}.  The initial goal leading to this phenomenon      
was to understand how intrinsic are the above-listed  mysterious properties to black holes and gravity.
       That is, the key question is:  \\
       
       {\it How unique are black holes?}  \\
       
       This question determines our the strategy which consists of the following steps.        
       
       \begin{itemize}
  \item 1. Carefully define the universality class of objects 
      of  high efficiency of information storage;
  \item 2. Construct calculable prototypes; 
  \item 3. Check if and under what circumstances they exhibit the black hole type features; 
   \item 4. Identify the underlying microscopic mechanisms;
    \item 5. Discover new associated phenomena;
     \item 6. Go back to black  holes and apply the gained  knowledge.
    
\end{itemize}    

        In what follows, we shall undergo the above steps with 
        a specific focus on the phenomenon of memory burden and its new manifestations.

      \subsection{Systems of high efficiency of information storage: Assisted gaplessness} 
    
     We shall proceed in the following steps.  
      
    First we introduce the universality class of systems 
      of high efficiency of information storage.
    This class was defined according to an universal 
    mechanism of ``assisted gaplesness" introduced in its bare essentials in  \cite{Dvali:2017nis, Dvali:2017ktv, Dvali:2018vvx, Dvali:2018xoc, Dvali:2018xpy, Dvali:2018tqi}.   This mechanism is responsible 
    for the reduction of the energy cost  of the 
    information storage within a given system. 
  It was shown to  generically lead to the memory burden effect \cite{Dvali:2018xpy}. 
          
  After defining this universality class, we shall narrow it down  by the requirement of a consistent QFT embedding  of the system. This requirement restricts the information storage capacity  imposing  the universal upper bounds on microstate degeneracy \cite{Dvali:2019jjw, Dvali:2019ulr, Dvali:2020wqi}.   
 This concept naturally introduces a class of objects, the so-called ``saturons" 
  \cite{Dvali:2020wqi},  that saturate the QFT bound on the information capacity.

Although, the memory burden effect is most prominent in saturons, it is shared by a wides class of efficient information storers. This is because all such systems employ the mechanism of the assisted gaplessness.  
   We therefore start our discussion  by describing this mechanism. 

      In general, information is stored in features of the 
      system that can be rearranged in various patterns.  
    Already at the intuitive level, one understands that 
    for a high efficiency of information storage what should count is the diversity of the available features  and 
  the effortlessness of their inscription and rearrangements. 
   For instance, it is easier to type text in ink on paper, rather than to curve it in stone.         
   
    In QFT language this  boils down to 
   the diversity of the excitable degrees of freedom
   and the smallness of the energy gaps required for 
   their excitations.   
   
   Adopting the terminology of \cite{Dvali:2018xpy}
    we shall call these degrees of freedom the ``memory modes" and shall introduce them as quantum oscillators
   with creation  and annihilation operators,  
  $ \hat{a}_j^{\dagger}, \hat{a}_j $, where 
  $j =1,2, ..., N_{\rm M}$ is the mode-label.  
     
   Memory modes can satisfy either Bose or Fermi  creation-annihilation algebra  with    
   $ [\hat{a}_i, \hat{a}_j^{\dagger} ]_{\pm} \, = \, \delta_{ij}$ and all other (anti)commutators zero. 
 
   The diversity of ``flavors" of the  memory modes,  $N_{\rm M}$, measures the richness of the available  information patterns, also called the ``memory patterns". 
    These patterns are created by the sequences of the
    occupation numbers and are represented via the ket-vectors in the  Fock space of states, 
  \begin{equation} \label{pattern}    
    \ket{\rm p} = \ket{n_1,n_2,...,n_{\rm M}} \,, 
   \end{equation} 
    where numbers $n_j \equiv  \bra{\rm p} \hat{n}_j \ket{\rm p}$  represents the eigenvalues of the corresponding 
    memory-mode number operators, 
  $\hat{n}_{j} \equiv \hat{a}_j^{\dagger} \hat{a}_j$. 
    The space formed by all possible memory patterns shall 
    be referred to as the ``memory space" \cite{Dvali:2018xpy}.

  Of course, a formation of a particular information pattern 
   costs energy which depends on the structure 
   of the Hamiltonian.   
     For example,  consider  a free Hamiltonian 
     \begin{eqnarray}       
    \label{Hfree} 
 \hat{H}_{\rm mem} &\equiv &    \sum_{j=1}^{N_{\rm M}}  \,  m_j  \,  \hat{n}_j \,.  
\end{eqnarray}
describing  a set of non-interacting memory modes with the
energy gaps $m_j$.   The energy 
cost of the pattern (\ref{pattern}) then is
          \begin{equation} \label{Pvac} 
E_{\rm p} =  \bra{p}\hat{H}_{\rm mem} \ket{p} \, = 
 \, \sum_{j=1}^{N_{\rm M}} \,  m_jn_j \,.      
\end{equation} 
  The measure of the efficiency of the information storage
 is given by the following two parameters:  \\
 
 {\it 1)}  the absolute energy cost of the patterns;  
 
 {\it 2)}  the density of the 
  pattern spectrum.  \\
  
   Obviously, the efficiency is higher with larger $N_{\rm M}$  and smaller $m_j$. 
 
  However, for ordinary systems, the information storage efficiency is rather poor. 
  For example, for a typical quantum oscillator originating 
  from a momentum-mode of a  relativistic field  localized in a box of size $R$, the characteristic  energy gaps satisfy $m_j \gtrsim 1/R$. 
   
    Now, the defining property of systems of high efficiency 
    of information storage is that they posses a mechanism for significantly reducing the energy gaps of the memory modes.  That is,  all such systems create an environment in which the memory modes become gapless, or nearly-gapless.

     The modes responsible for creating such an environment 
     shall be called the ``master modes"   \cite{Dvali:2018xpy}. 
      Their creation/annihilation operators 
     shall be denoted by  Greek symbols,  such as
      $\hat{\alpha}^{\dagger}, \hat{\alpha}, ...$.         
   These modes do not require a large diversity. 
   However,  they must posses the following features. 
   
   Firstly,  the master modes must be ``soft" (i.e.,  posses relatively small energy gaps / frequencies). This allows to reach the macroscopic occupation numbers  at relatively low energy cost.  In other words, the master modes 
   create a nearly-classical background field of low energy. 
   
   Secondly, the master modes  must interact with the memory modes in an attractive manner, meaning that 
the background of the master mode must lower  (``redshifts")  the effective  frequencies of the memory modes. 
In this way,  for certain critical occupation number 
$N$ of the master mode, the memory modes 
become (nearly)gapless.

     For reducing the mechanism of assisted gaplessness 
 to  its bare essentials, we shall use the following simple prototype Hamiltonian \cite{Dvali:2018xpy, Dvali:2018ytn, Dvali:2020wft}, 
 \begin{eqnarray}       
    \label{Hint} 
 \hat{H} \, &=& \, \hat{H}_{\rm ms} \, + \,  \hat{H}_{\rm mem} \,, \\ \nonumber 
 {\rm with}:~ 
 \hat{H}_{\rm ms} &\equiv& m_{\alpha}  \hat{n}_{\alpha} \,, \\ \nonumber   
\hat{H}_{\rm mem} &\equiv &  \left (1 -  \frac{
 \hat{n}_{\alpha}}{N} \right )^p  \sum_j m_j  
  \hat{n}_j \,.  
\end{eqnarray}
Here $p$ is a positive parameter, $N$ is a large number and 
$m_{\alpha}$ is the energy gap parameter of the 
 master mode with the number operator  $\hat{n}_{\alpha} \equiv \hat{\alpha}^{\dagger} \hat{\alpha}$.  This mode satisfies the  bosonic creation-annihilation algebra 
 with $[\hat{\alpha}, \hat{\alpha}^{\dagger} ]_- = \, 1$.

    In order to see how the Hamiltonian (\ref{Hint}) 
    leads to the assisted gaplessness,  let us consider the states in which the master mode is occupied to the critical value $n_{\alpha} = N $. 
    It is clear that in such a state, the energy gaps 
    of the memory modes collapse to zero and the 
    patterns (\ref{pattern}) become degenerate. 
   In other words, the master mode {\it assists}  the memory modes in becoming gapless.  
   
    In such  critical states, the memory space becomes  
    energetically flat.  All possible patterns (\ref{pattern}) 
   become promoted into a set of degenerate microstates.  The corresponding microstate entropy 
  is determined by the range of the memory-mode occupation numbers.   For example, if these numbers are uncorrelated and are individually bounded by $n_j \leqslant  d_j$, the 
  number of states is $n_{\rm st} = \prod_j (d_j + 1)$ with the corresponding microstate entropy, 
  \begin{equation} \label{Sd}
  S = \ln (n_{\rm st}) = \ln (\prod_j (d_j+1)) \,.
  \end{equation} 
   In particular, for uncorrelated fermionic memory modes, $d_j =1$,  which gives 
  $n_{\rm st} = 2^{N_{\rm M}}$ and   $S = N_{\rm M} \ln 2 $.

  However,  in many cases the occupation numbers of the memory modes are
   correlated. For example, a special mechanism 
   of  endowing an object  (e.g., a soliton) by 
   high efficiency of information storage  is via a spontaneous 
   breaking of some large symmetry  $G$ down to its subgroup $G'$ in the interior of the 
   object.  This mechanism was originally proposed 
   in \cite{Dvali:2019jjw} and further applied  to various systems \cite{Dvali:2019ulr, Dvali:2020wqi, Dvali:2021tez,  Dvali:2024hsb}.

   In these models, the memory modes 
   emerge as the Goldstone bosons of  spontaneously broken symmetry.  These modes are gapless 
   and are localized within the object that breaks the symmetry spontaneously.  In the exterior 
   vacuum the symmetry is restored.    
    Correspondingly, the memory space in determined by 
  the quotient space $G/G'$. 
   Later, we shall discuss such a setup in great details 
   and  shall use it for demonstrating the swift memory burden effect in soliton mergers.

   In general, up to log-factors, the microstate entropy 
  is typically given by the diversity of the memory modes, 
       \begin{equation} \label{SNm}
  S \sim  N_{\rm M} \,.
  \end{equation} 
    Next, we shall discuss the consistency bounds on this quantity.

  \subsection{Three incarnations of field theoretic bound on information storage capacity} 
  
The above discussion and the equation (\ref{SNm}) may create an impression  that the microstate entropy of the memory space is unlimited, as one can arbitrarily increase 
the diversity of the memory modes $N_{\rm M}$. This however is not true,
since the Hamiltonian (\ref{Hint}) must be viewed as 
an effective description emerging from 
a consistent QFT. 
The validity of the given QFT description imposes highly 
 non-trivial constraints on the parameters of the theory. 
  These constraints translate into  the constraints  on the parameters of the Hamiltonian (\ref{Hint}) and correspondingly into the bounds on the microstate entropy $S$. 
  
   The QFT-validity bounds on  $S$ were derived in \cite{Dvali:2019jjw, Dvali:2019ulr, Dvali:2020wqi}  where it was shown that  in $d$ space-time dimensions
  any object  localized within  a $d-2$-dimensional sphere of radius $R$ must satisfy the following upper bound  on the microstate entropy,   
   \begin{equation} \label{Area}
       S_{\rm max} \,  \sim  \,  (R \, f)^{d-2}
       \end{equation}   
where $f$ is the scale of spontaneous breaking of Poincare symmetry  by the object in question.  Notice that 
any localized macroscopic object breaks Poincare symmetry spontaneously and in each case the scale $f$ is 
unambiguously determined from the solution.  

In  \cite{Dvali:2019jjw, Dvali:2019ulr, Dvali:2020wqi}, it was also shown that the bound can be written in terms of the coupling $\alpha$ of the interaction that is responsible for the existence of the localized object,         
    \begin{equation} \label{Alpha}
       S_{\rm max} \,  = \, {1 \over \alpha} \,. 
       \end{equation}   
 In the above equation, the running coupling $\alpha$  has    
  to be evaluated at the scale set by the localization radius 
$R$.  
         
    The bounds  (\ref{Area}) and (\ref{Alpha})   
    mark the validity of a given QFT description. 
    Namely, it was shown that their violation  is correlated with the breakdown of loop-expansion 
    \cite{Dvali:2019jjw,Dvali:2019ulr} as well as with the violation of unitarity by a set of multi-particle scattering amplitudes \cite{Dvali:2020wqi}.

  In \cite{Dvali:2020wqi}, the objects saturating the above bounds were called  ``saturons".  
  As shown in the series of papers 
 \cite{Dvali:2019jjw, Dvali:2019ulr, Dvali:2020wqi,
Dvali:2021ooc, Dvali:2021jto, Dvali:2021rlf, Dvali:2021tez, Dvali:2021ofp, Dvali:2023qlk, Dvali:2024hsb}   
   saturons  reproduce all the previously-listed 
   ``mysterious" properties of a black hole with the mapping $f \rightarrow  M_P$. 
 
  In particular, one can immediately notice
 \cite{Dvali:2019jjw, Dvali:2019ulr, Dvali:2020wqi} 
   that the  Bekenstein-Hawking entropy of a black hole 
   (\ref{BekH}) represents a particular manifestation of  the area-law bound (\ref{Area}), since  the scale of Poincare-breaking by a black hole, regardless of its mass,  is $f= M_P$ (see below).  
 
  The equivalence between other saturons and black holes indicates that  they belong to the same universality class.  In particular, they all share the memory burden effect. 
    The memory burden effect in solitons and 
    its parallels with black holes has been studied in 
    \cite{Dvali:2021tez, Dvali:2024hsb}.

  In the present paper, we shall see that the same applies to the swift memory burden response.  In particular,  we shall demonstrate  later that this  phenomenon also takes place for saturated solitons. 
 
    Now, since the prototype Hamiltonians (\ref{Hint})  are assumed to  be effective descriptions of underlying  QFTs, 
    the bounds (\ref{Area}) and (\ref{Alpha}) must be
    taken into the account.    
    These bounds translate as the constraints on the parameters of the Hamiltonian   (\ref{Hint}) in the following way.   
    
    The bound (\ref{Alpha}) tells us that 
   the system described by  (\ref{Hint}) shall reach the 
   limit of information storage capacity when the 
   microstate  entropy  $S$ becomes equal 
   to the inverse coupling of the master mode, $\alpha_{\rm master} = 1/N$.  Taking into account (\ref{SNm}), this implies the bound on the diversity of the memory modes 
   in terms of the critical occupation number of the master mode, 
        \begin{equation} \label{MNbound}
    N_{\rm M} \lesssim N  \,. 
   \end{equation} 
         
  Next, for taking into account the bound (\ref{Area}), 
 we first notice that the size of the system is related with the 
 gap of the master mode as  $R \sim  m_{\alpha}^{-1}$. 
   Now assuming that the system is bounded by a $(d-2)$-sphere, 
   the scale of Poincare-breaking satisfies: 
   \begin{equation} \label{fPvalue}
    f^{d-2} = m^{d-2} N\, . 
   \end{equation} 
     Then,  the area bound (\ref{Area}) 
    translates as the following entropy bound, 
    \begin{equation}
     S_{\rm max}  \sim  \left(\frac{f}{m_{\alpha}} \right )^{d-2} \sim   N\,.  
    \end{equation} 
     Taking into account (\ref{SNm}), we again arrive to the 
     bound (\ref{MNbound}). 
     
     We thus see that, independently of the value of the  energy gap  of the master mode, $m_{\alpha}$, the system of   the efficient information capacity must satisfy (\ref{MNbound}).  
     
       In all the saturated QFT systems the relation 
       (\ref{MNbound}) has been observed explicitly.
     As already noticed in \cite{Dvali:2019jjw}, this implies 
     that the microstate entropy of a saturated system 
     cannot exceed the total occupation number of 
     constituent quanta. That is, in addition to 
     (\ref{Area}) and (\ref{Alpha}) there exist yet another 
     form of the QFT-bound on the microstate entropy \cite{Dvali:2019jjw}: 
        \begin{equation} \label{Number} 
     S_{\rm max} \,  \sim \,  N\,.  
    \end{equation} 

 In conclusion, the QFT-validity puts an universal upper  bound  on the microstate entropy which has at least three different physical meanings expressed 
 by the equations (\ref{Area}), (\ref{Alpha}) and 
 (\ref{Number}), respectively. 
  In the language of memory and master modes 
  they all imply the relation (\ref{MNbound}).

  However,    the following must be said. 
  The memory burden effect,  and in particular the swift response which is the focus of the present work,  does not require the saturation of the above bounds. 
 For its manifestation it suffices that the system has high efficiency  of information storage without having the maximal one. 
   In this sense, the memory burden effect extends beyond the  class of  maximal information capacity, which makes its manifestations wider spread.

    \subsection{Universality of information retainment} 
 Before moving to the proper memory burden effect, we 
 would like to focus on a first universal feature of systems 
of efficient information storage,  noticed in \cite{Dvali:2018xpy}. Namely, at initial stages of time evolution, such systems retain information internally.  

The reason is  the gaplessness of the memory modes.  
Indeed, the extraction of 
information requires a non-trivial time evolution of the 
memory modes.  However, in the critical state of efficient information storage, their frequencies are extremely 
suppressed.  This suppresses their time evolution.  
 In other words,  the state of efficient information storage 
 makes impossible a fast read-out of the information.  

 In other words, in order to make the information accessible, 
 the system must first move away from the critical point. 
   Achieving this via a quantum decay requires a macroscopic time.   

 This observation is of general physical importance, as 
 it reveals that the inability to  emit information at initial stages of the decay is not an exclusive property of a black hole. Rather it is an universal property of any device 
 that stores information efficiently.  Moreover, physics 
 behind this effect if fully exposed by the Hamiltonian 
 (\ref{Hint}).

 \subsection{Memory burden effect}

     We are now prepared to discuss the memory burden effect 
    \cite{Dvali:2018xpy, Dvali:2018ytn, 
 Dvali:2020wft}.  The essence of this phenomenon 
 is the following. 
 On top of the vacuum of the master mode, 
 $n_{\alpha} =0$, a loaded memory pattern,  would cost energy, given by (\ref{Pvac}). 
 This energy cost can be extremely high.

  In contrast,  in the critical state, $n_{\alpha} \, = \, N$,
  the same memory pattern costs a  (nearly)zero energy.  
  Of course, this happens at the expense of the energy 
  of the muster mode, which in the critical state 
  takes the value, 
            \begin{equation} \label{Ems} 
E_{\rm ms}  =  m_{\alpha} N \,.      
\end{equation} 

 We can say that the information-storage is energetically  efficient as long as the master mode's energy 
$E_{\rm ms}$ (\ref{Ems}) is less than the pattern's vacuum  energy $E_p$ (\ref{Pvac}).  
 
 In order to quantify the efficiency of the information storage,  
  it is useful to define the memory burden  parameter, 
             \begin{equation} \label{MuM} 
\mu \equiv \frac{E_{\rm ms}}{pE_p}  =
  \frac{m_{\alpha} N}{pE_P} \,,      
\end{equation} 
which measures the energy invested in the master mode 
versus the energy-cost of the pattern in the vacuum. The 
critical exponent of the gap function, $p$, is included for convenience. 
 
   The parameter $\mu$ also measures the memory burden back-reaction when the system is moved away from the critical point, $n_{\alpha} = N$.  
  For smaller $\mu$ the information-storage is more energy efficient and, correspondingly, the memory burden response is stronger.

 To be more precise,  the critical value is $\mu =1$. 
 For  $\mu \leqslant 1$, the minimum 
  of the energy is achieved for  the following occupation number of the master mode:
          \begin{equation} \label{nphiA}  
n_{\alpha} \, = \, N
\left (1 \, - \, \mu^{\frac{1}{p-1}} \right ) \,,   
\end{equation}   
  for which the energy of the system is given by 
          \begin{equation} \label{EnphiA}  
E \, = \, m_{\alpha} N 
 \left (1 \, - \,  \frac{p-1}{p} \mu^{\frac{1}{p-1}} \right ) \,.   
\end{equation}    
   This energy is less than the vacuum energy cost of the pattern (\ref{Pvac}).  Correspondingly,  beyond this point 
   the system resists to lowering the 
   occupation number of the muster mode.
  It will back-react  and block 
  any external factors  that attempt lowering  $n_{\alpha}$. 

\subsection{More general gap functions} 

  One can certainly consider the generic systems of the efficient information storage in which the assisted gaplessness 
 is achieved  via  the gap functions with more complicated dependences on the master and the memory modes,  
   \begin{eqnarray}       
    \label{HintG} 
\hat{H}_{\rm mem} &\equiv &    \sum_j \,  {\mathcal G_j}(\hat{\alpha}^{\dagger}, \hat{\alpha}) \, \hat{a}_j^{\dagger} \hat{a}_j \,  +  \\ \nonumber 
  &+&  \sum_{i,k}  \, {\mathcal G_{ik}}(\hat{\alpha}^{\dagger}, \hat{\alpha}) \,  \hat{a}_i^{\dagger} \hat{a}_k^{\dagger} \, 
+   \sum_{i,k} \,   {\mathcal G_{ik}}^{\dagger} (\hat{\alpha}^{\dagger}, \hat{\alpha}) \,  \hat{a}_i \hat{a}_k  \,.
\end{eqnarray}
   In this parameterization,  the  previously-considered memory mode Hamiltonian
   (\ref{Hint}) represents a particular case of (\ref{HintG}) 
  with ${\mathcal G_{j}} =  \left (1 \, - \,  \frac{\hat{n}_{\alpha}}{N} \right )^{p}m_j$ and ${\mathcal G_{ik}} = 0$.  
  
     This does not change the essence of the story. 
   First,    
    one needs to diagonalize the memory modes by the 
   proper Bogoluibov transformations,  
    \begin{eqnarray}       
    \label{Btrans} 
 \hat{a}_i \, &=& \,  u_{ij} \hat{b}_{j}  + v_{ij}^*\hat{b}_{j}^{\dagger}  \,,
\end{eqnarray}
 where $u_{ij}, v_{ij}$ are the Bogoliubov coefficients, which depend on the gap functions
 ${\mathcal G_{j}}$ and ${\mathcal G_{ik}}$. 
 Notice that only the memory modes are subjected to the Bogoliubov transformations, whereas the master modes 
 are treated as $c$-numbers that satisfy
   $\hat{\alpha}^{\dagger} \alpha  = n_{\alpha}$.   
 
 The resulting Hamiltonian of the Bogoliubov  
  memory modes has the form, 
     \begin{eqnarray}       
    \label{HintGB} 
\hat{H}_{\rm mem} &= &    \sum_j \,  {\mathcal G_j}^{B}(\hat{\alpha}^{\dagger}, \hat{\alpha}) \, \hat{b}_j^{\dagger} \hat{b}_j \,, 
\end{eqnarray}
 where ${\mathcal G_j}^{B}$ is the diagonalized gap function.  
 
    Now,  the statement that the system has a high efficiency of information storage, implies that the gap function 
    for the Bogoliubov modes ${\mathcal G_j}^{B}$ reaches zero for certain  critical occupation numbers of the master modes.  
   
    In other words,  the generic system of efficient 
  information storage (\ref{HintG}) is defined 
  by the feature that the assisted gaplessness 
  is exhibited by the Bogoluibov modes which diagonalize the Hamiltonian around the critical point.

   If such a critical point is non-existent, the system is 
 outside of our interest.          
     On the other hand, for a system that possesses  
 the state of the assisted gaplessness,  the memory 
 burden effect is imminent and goes as discussed previously.   
 That is, by a proper Bogoliubov transformation,
 a generic system of the efficient information storage 
 (\ref{HintG}), near the critical point of the assisted gaplessness 
  effectively reduces to the prototype Hamiltonian  (\ref{Hint}), modulo the number of the master modes. 
 
 The reason for the robustness of the memory burden effect in such systems is that  the point of the assisted gaplessness is a type of a quantum critical point that exhibits a universal behaviour.  
   We shall demonstrate  these features later when we consider some explicit examples of the Bogoliubov diagonalizations of the memory modes.     
  At the moment, for the sake of making the discussion maximally transparent,  we stick to the simplified Hamiltonians of the type (\ref{Hint}).  These 
  Hamiltonians fully capture the essence of the effect 
 while avoiding the additional unessential technicalities.

 \subsection{Entanglement of memory modes}
 
  As discussed, in the state of the assisted gaplessness, 
 the memory space is highly degenerate in energy. 
 Therefore, the system can exist in a state of superposition 
 of various memory patterns \cite{Dvali:2018xpy}, 
 \begin{equation} \label{EntPattern}
   \ket{{\rm mem}} = \sum_p \, c_p \, \ket{p} = 
   \sum \, c_{n_1,..n_{N_{\rm M}}} \ket{n_1,n_2, ...,n_{\rm M}} \,, 
  \end{equation}  
  where  $c$-s are coefficients and the sum is taken over the entire memory space. Of course, in such a state the memory modes are entangled.  
 In a typical highly entangled state, the number of contributing basis patterns will be of the order of the dimensionality of the memory space.  
 
   The system can also be in very rare EPR-type states. 
   For example, for qubit-type memory modes with two choices of the occupation numbers, $n_j = 0, 1$,  such a state is,    
    \begin{equation} \label{EPR} 
   \ket{{\rm mem}} = \frac{1}{\sqrt{2}} ( \ket{0,0, ...,0}  +   \ket{1,1, ...,1} )  \,. 
  \end{equation}  
   
   A curious thing about the entangled memory states
is that in the state of the assisted gaplessness they 
are the eigenstates of the Hamiltonian 
(\ref{Hint}), whereas 
away from it, are not.   For example, the two basis states 
entering in  (\ref{EPR}) are degenerate 
for $n_{\alpha} = N$ and are maximally split 
for $n_{\alpha} = 0$.   
Correspondingly, these  two basis states entering 
the superposition will contribute very differently into the memory burden effect. 

 The entanglement of the memory modes does not 
 change the very existence of the memory burden effect.  
 However,  it distributes the weight of the burden among different basic patterns entering the superposition.
 Due to this, it makes the dynamics of  
 quantum information processing reacher.
   From the point of view of the basis defined by the patterns (\ref{pattern}), the information stored in the entangled memory modes  
  (\ref{EntPattern}) appears to be  ``shuffled up".  
  In particular, in imprints in the black hole mergers, 
the entanglement of the memory modes will manifest 
itself in the contributions of the higher order correlators.

  \section{Dynamics of memory burden: gradual 
   versus swift} 
  
   We shall now discuss the dynamics of the memory burden 
 effect using some explicit examples of the disturbances of the system. 
     These examples shall suffice for understanding 
     the universal nature of the phenomenon 
     as well as the differences between the two  regimes of the memory burden: 
     gradual versus swift.
     The regime is determined by the nature of  the external stimulus.  
     
     A strong classical perturbation, such as a merger with another black hole,  is met with a swift memory burden response.  The time-scale of the effect is determined by the time-scale of the classical perturbation experienced by a black hole. 
 
On the other hand, in the absence of the external influence, the memory burden sets in gradually, in form of a back-reaction 
against the slow quantum decay of the system.

   Following the original work   \cite{Dvali:2018xpy, Dvali:2018ytn, Dvali:2020wft}, this can be illustrated by a simple interaction hamiltonian that  enables the change of the occupation number of the master mode. 
  
    In particular,  we can allow the transition of the master mode into an external mode $\hat{\beta}$. 
   This can describe the evolution of the system 
 in  different physical situations.
 
  In particular, 
 quanta $\hat{\beta}$ can  impersonate  the  free asymptotic quanta to which the system decays.  For example, such can be  the quanta in Hawking radiation of a black hole.  
 
  Alternatively,  $\hat{\beta}$ can act as  a coherent mode of an external classical field  
 to which the system interacts. For example, this role 
 can be played by the coherent modes of the gravitational field excited in black holes mergers.  
  In particular,  the modes can describe an outgoing gravitational radiation  as well as the quasi-normal excitations of the black hole. 
 
  We shall illustrate the effect for two types of interaction Hamiltonians, with linear and non-linear interactions of 
  the master mode  $\hat{\alpha}$ with the external degree of freedom 
  $\hat{\beta}$.

   \subsubsection{Memory burden for linear mixing}

  First, we discuss the transition between the 
  two sectors induced via the following simplest mixing terms  
   in the interaction Hamiltonian \cite{Dvali:2018xpy, Dvali:2018ytn, Dvali:2020wft},  
         \begin{equation} \label{Hextra} 
\hat{H}_{int} \, = \,  
 \tilde{m} \, \hat{\beta}^{\dagger} \hat{\alpha} \, +  \,
\tilde{m}^* \, \hat{\alpha}^{\dagger} \hat{\beta} \,   
\, +  \, \omega_{\beta} \hat{n}_{\beta} \,,    
 \end{equation} 
 where $\hat{n}_{\beta} \equiv \hat{\beta}^{\dagger}\hat{\beta}$ is the $\beta$-mode number 	operator and 
 $\tilde{m}$ is a complex parameter of dimensionality of
 energy. 
  For maximizing the effect, we take the 
   energy gaps of the two modes to be degenerate 
   $m_{\alpha} = \omega_{\beta}$.  
      
  As we shall explain later, this choice is also dynamically-justified. This is because  in realistic situations, in which the  external modes $\hat{\beta}$ come from a continuous 
    spectrum, the system always transits into the resonant 
    mode of the nearest frequency.

   Following  \cite{Dvali:2018xpy, Dvali:2020wft}, 
 it is easy to see that the information pattern of the memory modes  dramatically affects the time-evolution of the 
system. 

  For  an empty pattern, $E_p =0$,  starting  from initial state $n_{\alpha} = N$  and  $n_{\beta} =0$,  the  occupation numbers   evolve in time  as, 
   \begin{equation} \label{Nsin} 
\frac{n_{\alpha}(t)}{N}\, = \,   \cos^2\left (|\tilde{m}|
t \right ) \,, ~~\, 
\frac{n_{\beta}(t)}{N}\,= \,\sin^2\left (|\tilde{m}|
t \right ) \,. 
 \end{equation}
   Thus, the master mode 
 gets fully depleted within the time $t \sim \pi/(2|\tilde{m}|)$.

  However, for  $ \frac{|\tilde{m}|}{m_{\alpha}} \, \mu  \ll 1$,  
  the story is 
  dramatically  different.  The relative change of the master mode $\Delta n_{\alpha} \, \equiv  \, N \, - \, 
 n_{\alpha}$,  induces a back reaction 
 from the memory pattern that  shifts the gap of the 
 $\hat{\alpha}$-mode by, 
           \begin{equation} \label{gap} 
\delta m_{\alpha} = p N^{-1} \left ( \frac{\Delta n_{\alpha}}{N} 
\right )^{p-1} E_P \, = \frac{m_{\alpha}}{\mu}  \left ( \frac{\Delta n_{\alpha}}{N} 
\right )^{p-1}.    
 \end{equation}
 
 This induces the level-splitting
 between $\hat{\alpha}$ and $\hat{\beta}$ modes  and effectively blocks the time evolution once the change of occupation number reaches the following critical value 
           \begin{equation} \label{critD} 
 \frac{\Delta n_{\alpha}}{N} \, \simeq 
\left ( \frac{|\tilde{m}|}{m_{\alpha}} \, \mu \right )^{\frac{1}{p-1}} \,.      
 \end{equation}
  Taking into account (\ref{Nsin}), it is easy to estimate the corresponding time \cite{Dvali:2025ktz}, 
              \begin{equation} \label{MBtime} 
t_M \, \simeq 
\frac{1}{|\tilde{m}|}\left ( \frac{|\tilde{m}|}{m_{\alpha}} \, \mu \right )^{\frac{1}{2(p-1)}} \,.    
 \end{equation}
  This expression gives the time-scale required for the system to enter the memory burden regime. 
  
   The above analysis can be easily generalized to the 
   transition into multiple external species. 
  For this, we allow the master mode to transit 
  to $N_{sp}$ copies of the $\hat{\beta}$-species 
  $\hat{\beta}_k$, with $k = 1,2,...,N_{sp}$. 
  The interaction Hamiltonian is,  
           \begin{equation} \label{HextraN} 
\hat{H}_{int} \, = \,  
 \tilde{m} \, \hat{\beta}_i^{\dagger} \hat{\alpha} \, +  \,
\tilde{m}^* \, \hat{\alpha}^{\dagger} \hat{\beta}_i \,   
\, +  \, \omega_{\beta} \sum_{i=1}^{N_{sp}} \hat{n}_{\beta_i} \,.    
 \end{equation} 
  For definiteness, we again assume that the modes are initially-degenerate, 
 $\omega_{\beta} = m_{\alpha}$.  
 Thus, the master mode effectively mixes with a single 
 external mode defined as, 
            \begin{equation} \label{Nbeta} 
\hat{\beta} \equiv  \frac{1}{\sqrt{N_{sp}}} \sum_{i=1}^{N_{sp}} \hat{\beta}_i \,,    
 \end{equation} 
but with  an enhanced mixing term, 
$\tilde{m}\sqrt{N_{sp}}$. Thus, the story is equivalent 
to the previous example with a single $\hat{\beta}$-mode, modulo
the rescaling $\tilde{m} \,  \rightarrow  \, \tilde{m}\sqrt{N_{sp}}$.  This immediately gives that with 
$N_{sp}$ external species the transition time to the memory burden is changed as 
             \begin{equation} \label{MBtimeN} 
t_M^{(N_{sp})} \, \simeq N_{sp}^{\frac{3-2p}{4(p-1)}}
t_M \,.    
 \end{equation}

Of course, in systems such as a black hole,
 due to the available phase space, the decay is not fully oscillatory (see below).  This however is no obstacle, since  as it is clear from 
(\ref{MBtime}) the  swift memory burden sets in way before a would-be  oscillation period is elapsed.  As we shall discuss, the system keeps  scanning over a continuum of the
 $\hat{\beta}$-modes with resonant frequences. 
  Therefore, already the simplest single-mode oscillatory 
  system correctly captures the onset of the effect.

 \subsubsection{Memory burden for non-linear evolution}

  Let us now discuss the memory burden effect 
  for a non-linear transition  
  between $\hat{\alpha}$ and $\hat{\beta}$ modes.  
  For definiteness,  we choose the interaction Hamiltonian in the  following form, 
    
           \begin{equation} \label{Hextra2} 
\hat{H}_{int} \, = \,  
 \frac{\tilde{m}}{2N} \, (\hat{\beta}^{\dagger})^2 (\hat{\alpha})^2 \, +  \,
\frac{\tilde{m}^*}{2N} \, (\hat{\alpha}^{\dagger})^2 (\hat{\beta})^2 \,   
\, +  \, \omega_{\beta} \hat{n}_{\beta} \,,    
 \end{equation} 
 where for convenience we have normalized
 the mixing terms via the universal coupling $1/N$.   
   With this Hamiltonian we shall study the transition of 
 the master mode  $\hat{\alpha}$ into the
 external modes $\hat{\beta}$ and see how the 
 transition  is influenced by the memory burden effect. 
  
    In our initial state the master mode is occupied 
    to a critical value $n_{\alpha} = N$, whereas the 
    external mode is in the vacuum, $n_{\beta} = 0$. 
    
     We  first study the transition for the empty memory 
     pattern, $E_p =0$ ($\mu = \infty$). 
   The initial stages of the depletion can be analysed using  the Bogoliubov approximation, in which the 
   creation and annihilation operators for the muster 
   modes are replaced by the $c$-numbers 
   $\hat{\alpha}^{\dagger} =  \hat{\alpha}  =  \sqrt{n_{\alpha} }$, 
   where the phases are assumed to be absorbed in the phase  of $\tilde{m}$. 
     Up to $1/N$-corrections, the  effective  Hamiltonian 
     of the $\hat{\beta}$-modes takes the form, 
                \begin{equation} \label{HextraEff} 
\hat{H}_{\beta} \, = \, \omega_{\beta} \hat{\beta}^{\dagger} \hat{\beta} \,  +  \,  
 \frac{\tilde{m}}{2}\, (\hat{\beta})^2  \, +  \,
\frac{\tilde{m}^*}{2}  \, (\hat{\beta}^{\dagger})^2  \,.    
 \end{equation} 
   Without loss of generality, we assume that $\tilde{m}$ is real and positive. 
   The above Hamiltonian is diagonalized by the following Bogoliubov transformations, 
                   \begin{equation} \label{BogoliubovT} 
 \hat{b}  = u  \hat{\beta} + v \hat{\beta}^{\dagger}  \,,    
 \end{equation} 
  with 
                   \begin{equation} \label{BUandV} 
 u^2 \,  = \, \frac{1}{2} \left ( 1 + \frac{1}{\sqrt{1 - \lambda}}\right)\, ~~ v^2 \, =  \, \frac{1}{2} \left ( - 1 + \frac{1}{\sqrt{1 - \lambda}}\right)\,,    
 \end{equation} 
  where $\lambda \equiv \frac{\tilde{m}^2}{\omega_{\beta}^2} $.
      The diagonalized free Hamiltonian of the Bogoliubov 
      modes has the form, 
                   \begin{equation} \label{HdiagB} 
\hat{H}_{b} \, = \, \omega_{\beta}
\sqrt{1- \lambda} \, ~
 \hat{b}^{\dagger} \hat{b}   \,.    
 \end{equation} 
    The occupation number of $\hat{\beta}$-particles 
    is given by the Bogoliubov coefficient, 
     \begin{equation}  \label{nBeta}
    \Delta n_{\alpha} =  n_{\beta} \, = \, v^2 \, =  \, \frac{1}{2} \left ( - 1 + \frac{1}{\sqrt{1 - \lambda}}\right) \,.  
     \end{equation} 
     However, the above expression does not take into account the back-reaction from the depleted quanta. 
        This back-reaction corrects the depletion coefficient non-trivially.   The back-reaction  is taken 
      into the account via the constraint: 
   \begin{equation}  \label{NaNbN}
  \hat{n}_{\beta} = N - \hat{n}_{\alpha} = \Delta n_{\alpha} \,, 
   \end{equation}     
which  after the depletion of  the master mode into $
\Delta n_{\alpha} = n_{\beta}$ quanta, corrects the effective Hamiltonian  (\ref{HextraEff}) of the 
 $\hat{\beta}$-modes as, 
        \begin{eqnarray}                   
                 \label{HextraEffC} 
\hat{H}_{\beta} \, &=& \, \left ( \omega_{\beta} +  \frac{m_{\alpha}}{\mu}\left  (\frac{n_{\beta}}{N} \right )^{p-1} \right)   \, \hat{\beta}^{\dagger} \hat{\beta} \,  +  \, 
\\ \nonumber  
 &+ & \frac{\tilde{m}}{2}  \left (1-\frac{n_{\beta}}{N} \right )\, (\hat{\beta})^2  \, +  \,
\frac{\tilde{m}^*}{2} \left (1-\frac{n_{\beta}}{N} \right ) \, (\hat{\beta}^{\dagger})^2  \,.    
 \end{eqnarray}  
  Correspondingly, the factor $\lambda$ in the Bogoliubov 
  coefficient is corrected as, 
      \begin{equation}  \label{lambdaC}
  \sqrt{\lambda} \,  = \,  \frac{\tilde{m}}{\omega_{\beta}}
  \frac{N-n_{\beta}}{N} 
  \frac{1}{ 1+  \frac{m_{\alpha}}{\mu \, \omega_{\beta}}\left  (\frac{n_{\beta}}{N} \right )^{p-1} }  \,. 
\end{equation}     
 From  (\ref{nBeta}) it is clear that the depletion of the master mode into $\hat{\beta}$-modes  
 gets tamed after the number of depleted quanta 
 reaches the critical value, 
     \begin{equation}  \label{lambdaC}
  n_{\beta} \,  \simeq \, N\, \left (\frac{\omega_{\beta} }{m_{\alpha}} \mu \right )^{\frac{1}{p-1}} \,. 
   \end{equation} 
   
   The above analysis is straightforwardly generalizable to 
   the depletion  to  $N_{sp}$  species  $\hat{\beta}_l$,  
 where $l= 1,2,...,N_{sp}$ is the species 
 label.   In this case, the  Hamiltonian  (\ref{Hextra2}) 
  gets replaced by, 
              \begin{equation} \label{HextraSP} 
\hat{H}_{int} \, = \,  
 \frac{\tilde{m}}{N} \, (\hat{\beta}_l^{\dagger})^2 (\hat{\alpha})^2 \, +  \,
\frac{\tilde{m}^*}{N} \, (\hat{\alpha}^{\dagger})^2 (\hat{\beta}_l)^2 \,   
\, +  \, \omega_{\beta}  \hat{\beta}_l^{\dagger} \hat{\beta}_l \,,    
 \end{equation} 
where the summation over the species index $l$ is assumed. 
  
  The initial depletion becomes $N_{sp}$ times more efficient with the total number of the depleted quanta now given by,  $\Delta n_{\alpha} = N_{sp} n_{b}$.   This gives the relation,   
       \begin{equation}  \label{nBetaSP}
    \Delta n_{\alpha} =  \sum_j n_{\beta_j} \, =  \, N_{sp} \, \frac{1}{2} \left ( - 1 + \frac{1}{\sqrt{1 - \lambda}}\right) \,, 
     \end{equation} 
  where the factor $\lambda$ in the Bogoliubov 
  coefficient is correspondingly changed as, 
       \begin{equation}  \label{lambdaCSP}
  \sqrt{\lambda} \,  = \,  \frac{\tilde{m}}{\omega_{\beta}}
  \frac{N-\Delta n_{\alpha}}{N} 
  \frac{1}{ 1+  \frac{m_{\alpha}}{\mu \, \omega_{\beta}}\left  (\frac{\Delta n_{\alpha}}{N} \right )^{p-1} }  \,. 
   \end{equation}  
   The equation  (\ref{nBetaSP}) has to be solved 
   self-consistently for  $\Delta n_{\alpha}$. 
    It is clear that for more species the memory burden 
    phase is reached faster. 
   
   \subsection{Swift versus gradual  memory  burden} 
   
   Memory burden effect  universally resists to any 
   departure of the  system from the critical state 
 of assisted gaplessness. 
   However,  such departures can take place under 
   different regimes. 
    In particular, one can distinguish the following two extreme  cases:   
    
    \begin{itemize}
  \item A gradual departure due to a slow quantum 
    evolution; 
  \item A fast coherent/classical evolution. 
  
\end{itemize}
        
      The both regimes can be realized by generic 
  interaction Hamiltonians that describe transitions 
  between the master mode and the external fields. 
  One and the same Hamiltonian can produce 
  either of the regimes depending on the nature 
  of perturbation.  
  
 This can be clearly understood from the examples considered above.  
  Both Hamiltonians, (\ref{Hextra}) and (\ref{Hextra2}), can describe 
 equally well the process of slow quantum decay  of the master mode as 
 well the coherent classical transitions between the master mode and the external fields. 
 
  For concreteness, let us focus on the   
  the interaction Hamiltonian  (\ref{Hextra2}). 
 We have already studied the memory burden effect 
 in the regime of  gradual quantum depletion of the master mode using the Bogoliubov method.   
 Let us now analyse the coherent transition between
 the same modes.  That is,  we shall be interested in the
 time evolution of the master mode  
 $\hat{\alpha}$ into the mode $\hat{\beta}$ that can also be described coherently.

 The master mode is already macroscopically 
 occupied. The coherence of $\hat{\beta}$ implies that 
 we can use Bogolibov approximation also for this mode.
  That is, we replace all number operators by 
  $c$-numbers and accommodate the constraint
  (\ref{NaNbN}) by the following parameterization,  
   \begin{eqnarray} 
  \label{thetaAB}
   \hat{\alpha}^{\dagger} &=&  \hat{\alpha}  =  \sqrt{n_{\alpha}}
  \equiv  \cos(\theta) \sqrt{N}  \\ \nonumber 
   \hat{\beta}^{\dagger} &=& - \hat{\beta}  =  i \sqrt{n_{\beta}}
 \equiv i \sin(\theta) \sqrt{N} \,,
  \end{eqnarray}
   where $\theta$ is an  angular parameter 
   satisfying $0 \leqslant \theta \leqslant 2\pi$. 
  For convenience,  we have introduced the relative factor $i$ which accommodates the sign dictated by the 
  minimization of  mixing term. 
    In this parameterization, the full Hamiltonian becomes,  
       \begin{eqnarray}                   
                 \label{Htheta} 
\hat{H} \, &=& \, 
Nm_{\alpha} + (\sin\theta)^{2p} \, E_p \,  +  
\\ \nonumber  
 &+ &  N (\omega_{\beta} - m_{\alpha}) \, \sin^{2}\theta \, -
  N \tilde{m} \cos^2\theta \sin^2\theta \,.    
 \end{eqnarray} 
    The minimum of energy is achieved for, 
     \begin{equation}  \label{SinMin}
    \sin^2\theta = \frac{m_{\alpha} - \omega_{\beta} + \tilde{m}}{2\tilde{m} \left ( 1 +  \frac{m_{\alpha}}{2 \tilde{m} \mu}  (\sin\theta)^{2(p-2)} \right)} \,.
    \end{equation}    
     It is instructive to evaluate this expression for  
     $m_{\alpha} = \omega_{\beta}$.   Then, for the 
     empty memory pattern,  $E_p=0$, 
    which also implies $\mu = \infty$,   
     we have 
     $\sin^2\theta = 1/2$. 
     
      On the other hand, if the pattern is heavily loaded, 
    $\mu \ll 1$ (more precisely if $\mu \ll \frac{m_{\alpha}}{2\tilde{m}}$),  
     the memory burden is strong 
    and the minimum is achieved for, 
       \begin{equation} \label{minimum} 
    \sin^2\theta \, \simeq \,  \left (\frac{\tilde{m} N}{pE_p} \right)^{\frac{1}{p-1}} \, = \,  \left (\frac{\tilde{m}}{m_{\alpha}} \mu \right)^{\frac{1}{p-1}} \,.
    \end{equation}
    Correspondingly, beyond this point,  any coherent evolution of the system with be  swiftly affected by the memory burden effect.

   \subsection{Null memory burden surface} 
 
     For certain Hamiltonians  one can define 
    the notion of a `` null memory burden surface". 
   This term refers to a trajectory in the Hilbert space 
   along which the memory modes 
   stay gapless.  If the time-evolution 
   of the system coincides with such a trajectory, 
   the system shall evolve without experiencing the memory burden.  
    
    The necessary condition is the vanishing of the gap function of the memory modes  ${\mathcal G}(\hat{\alpha}^\dagger, \hat{\alpha} )$
 throughout  the entire trajectory.
   For this, the following conditions must be satisfied. 
   
    First, the zeros of the mode-function must be continuously  degenerate for a set of the expectation values of  the master modes.  The null memory burden surface, 
   represents a manifold in the field space defined 
   by the  condition, 
                       \begin{equation} \label{GapFunction} 
 {\mathcal G}(\hat{\alpha}^\dagger, \hat{\alpha} ) = 0 \,.     
 \end{equation}

      A simple
  example is provided  by a mode function
 which depends on the  two species of the muster 
    modes  $\hat{\alpha}^{\dagger},  \hat{\alpha}$ 
    and $\hat{\beta}^{\dagger},  \hat{\beta}$  in the following 
    manner, 
                        \begin{equation} \label{GapFunctionN} 
 {\mathcal G}(\hat{\alpha}^\dagger, \hat{\alpha} ) = 
 \left ( 1 - 
\frac{\hat{n}_{\alpha} + \hat{n}_{\beta}}{N}  \right )^p\,.      
 \end{equation}   
   The corresponding Hamiltonian of the memory modes 
 is, 
                   \begin{equation} \label{nullH} 
\hat{H}_{\rm mem}  \, = \,  \left ( 1 - 
\frac{\hat{n}_{\alpha} + \hat{n}_{\beta}}{N}  \right )^p \sum_j m_j \hat{n}_j \,.     
 \end{equation} 
 The null memory burden surface is determined by the condition, 
  \begin{equation} \label{ABtoN}
{n}_{\alpha} + {n}_{\beta} = {N} \,.
  \end{equation} 
In order for the system to evolve on the null memory burden surface, it is necessary that the time-evolution generated  by the rest of the Hamiltonian  respects the constraint  (\ref{ABtoN}).  This can be achieved in very special situations. 
 
  For instance, this is the case for 
the interaction Hamiltonians of the type (\ref{Hextra}) 
and (\ref{Hextra2}), since they both respect the constraint
(\ref{ABtoN}). 
   However,  such  a  time-evolution  can only 
   be achieved for very specific Hamiltonians 
   for which the zeros of the mode function 
    represent the integral of motion.  
    
   In general, it is impossible for the system to 
 evolve along the null memory burden surface under arbitrary perturbations. 
  This would require that zero of the gap function 
  is enforced by 
  the symmetry of the entire Hamiltonian. 
  In a realistic QFT system, this would mean that 
  the memory modes are identically gapless 
  throughout the available Hilbert space, implying that 
  the system is trivial. 
   
      In order to understand this, consider the following 
     construction. We can confine the system to 
     null memory burden surface by imposing
     the condition (\ref{GapFunction}) as the Hamiltonian constraint
                      \begin{equation} \label{nullH} 
\hat{H}  \, = \, 
{\mathcal G}(\hat{\alpha}^\dagger, \hat{\alpha} )  \sum_j m_j \hat{n}_j \,  +   \,  X {\mathcal G}(\hat{\alpha}^\dagger, \hat{\alpha} )\,,....\,,     
 \end{equation} 
  where  $X$ is a Lagrange multiplier that ensures 
(\ref{GapFunction}).  
 However, such a system is trivial in the sense that it exhibits no dynamics of information processing; 
 the memory modes are gapless over the entire Hilbert 
 space. In such memory modes, the information can neither be stored nor retrieved. 
  In other words, the Hilbert scape splits into the superselection sectors according to the memory patterns carried by them. 
 
 Such systems are not interesting from the point of 
 view of the present discussion.    
   Indeed, the very concept of the assisted gaplessness 
   that ensures the efficient information storage 
   is based on the premise that the memory modes 
    become gapless only around very special critical states, 
    which can be reached or abandoned by time-evolution.

    Correspondingly, in  a QFT  
   employing the mechanism of the assisted gaplessness, the generic perturbations will take the system away from the null memory burden surface. 
    This will  activate the memory burden. 
    
   Therefore, evolution of a non-trivial  system on a null memory burden surface can only take place as 
   a temporary coincidence for certain special trajectories. 
   In particular, this cannot be the case for the time-evolution 
   that affects the memory space. 
    For example, if the system's information storage capacity 
   decreases during any evolution, the memory burden is 
   inevitable.  In particular, this is the case when the system's 
   memory capacity decreases due to a decay into 
   the external quanta. Example of such decay is 
   provided by Hawking evaporation of a black hole.  
     For such processes the gradual memory burden is imminent. 
     
   On the other hand, for other types of perturbations 
   the accidental evolution along the null memory burden surface cannot be excluded.  
    We come back to this question in the next section when we discuss the  swift memory burden effect in black holes.

    \section{Swift activation of the 
  memory burden in black hole mergers} 
  
 The previous applications  of the memory burden in black holes were due to the burden that is induced by their decay via the Hawking radiation. This effect is especially important 
 for PBH ~\cite{Zeldovich:1967lct, Hawking:1971ei,Carr:1974nx,Chapline:1975ojl,Carr:1975qj},
 which can be much lighter than the astrophysical ones.   
 Such memory-burdened black holes, which become long-lived, can have important cosmological implications, such as the opening up a new window for PBH  dark matter 
  \cite{Dvali:2020wft, Alexandre:2024nuo, Dvali:2024hsb},
  which was later reanalysed by taking into 
  account the smooth entrance into the burden 
  phase \cite{Dvali:2025ktz}.
  Various implications of the burdened 
  PBH for gravitational waves \cite{MBGW1, MBGW2, MBGW3, MBGW4, MBGW5, MBGW6, MBGW7,
MBGW8, MBGW9, MBGW10, MBGW11, MBGW12}, 
for sources of high energy particles
\cite{MBHP1, MBHP2, MBHP3, MBHP4, MBHP5} 
 and mergers \cite{MBmerger1, MBmerger2}
 have been discussed.  An incomplete list of references investigating various other  aspects can be found in 
 \cite{MB1, MB2, MB3, MB4, MB5, MB6, MB7, MB8, MB9, MB10, MB11, MB12, MB13, MB14, MB15, MB16, MB17, MB18, MB19, Oleg, AnaGiaMichael}.     
     
   The manifestation of the memory burden effect that we are proposing in the present paper is very different.  Our point is that the memory burden can be swiftly activated 
   in the mergers of arbitrarily large black holes, 
  including the supermassive ones. The influence of the memory burden on such mergers  can have macroscopic and  potentially-observable  effects.  
  
   The presence of a swift memory burden effect is independent of the age of a black hole.  That is, even  if a black hole 
  is at the early stages of its existence and therefore not experiencing the gradual memory burden effect due to a back-reaction from the Hawking decay, the burden 
  shall get activated as a response to the external classical 
  disturbance.  Such a disturbance can come, for example, 
   from a merger with another black hole 
  or a star.  
    
    Our estimates show that the standard dynamics of merger, that ignores the memory burden back-reaction,  
     can be affected by order-one effects. 
   It is important to understand that, although the origin of the memory burden is quantum,  we are talking about 
   the classically-observable imprints. 
   
   The strength of the imprint is measured by 
   the memory burden parameter, $\mu$, which depends on 
   the fraction of the memory-capacity actually used by a 
   black hole information pattern. 
   That is,  $\mu$ is determined by the number and type of the actualized memory modes required by the 
  black holes's information load. 
   This can be estimated based on our current knowledge 
   of the black hole formation. 
   
       We shall structure our discussion in the following way. 
  First, we  discuss the specifics of the mechanism of the assisted  gaplessness in black holes.  In particular, we 
  identify the origin of the memory and the master modes and count their diversity.   Next, we map these specific features on the  parameters of the prototype Hamiltonians and study the effect of swift  memory burden.  We derive some key formulas and use them for estimating the swift memory burden  effect in astrophysical black holes and in PBH.

   \subsection{Specifics of black holes} 
 
 In order to apply the swift memory burden effect 
 to black holes, we first need to set straight the nature of relevant degrees of freedom and identify their main  characteristics.  This will enable us to  reduce the mechanism of black hole 
  information storage to its bare essentials and 
  map it on a prototype calculable model of maximal capacity 
  of memory storage.  \\
  
   We shall rely on three pillars:  \\
   
   {\it 1)}  The universal nature of the memory burden phenomenon
  \cite{Dvali:2018xpy, Dvali:2018ytn, Dvali:2020wft,  Dvali:2021tez, Dvali:2024hsb} in  
systems with assisted gaplessness \cite{Dvali:2017nis, Dvali:2017ktv, Dvali:2018vvx, Dvali:2018xpy} \cite{Dvali:2018tqi};  \\

   {\it 2)} The knowledge gained 
   from studying a wide variety of the prototype 
   systems  \cite{Dvali:2018xpy, Dvali:2018ytn, Dvali:2020wft,  Dvali:2021tez, Dvali:2024hsb}; \\
   
   {\it 3)}   The indications from the microscopic theory of black hole's $N$-portrait \cite{Dvali:2011aa, Dvali:2012en, Dvali:2013eja, Dvali:2012en, Dvali:2015ywa,  Dvali:2015wca,  Averin:2016hhm}. \\
  
   The first task is the identification of black hole's memory modes.  
   
 \subsubsection{Memory modes of a black hole}  
 
   We first define the black hole memory modes and 
   then justify the statement from various angles. \\
   
  {\it  The black hole memory modes are the gapless modes that 
  can be labeled  by the eigenvalues of the angular momentum (spherical harmonics). Such modes are ``deposited" by the graviton as well as by other species of quantum fields. 
   Correspondingly,  in addition to angular momentum, 
   the memory modes also carry the species label.
  The angular momentum can take arbitrary values 
  all the way to the UV-cutoff of the theory.   
   
  One can say that the black hole memory modes 
  represent particles of arbitrarily short wavelengths
  ``orbiting"  the black hole. 
   The unusual thing is that, despite their short 
   wavelengths (high momenta) in the black hole 
   vacuum these particles  have zero energies.  This is because of these modes are strongly off-shell as compared to their asymptotic counterparts. This is due to the assisted 
   gaplessness of the gravitational field. 
   
   However, when the black hole is subjected to a perturbation, it changes the gap of the memory modes and activates the swift memory burden effect.
    } \\

 One line of reasoning  supporting the above 
 description of black hole memory modes is based on 
 indications from  the microscopic theory of  
black hole's quantum $N$-portrait \cite{Dvali:2011aa}. 
    According to this theory, a  black hole represents a saturated coherent state (or a condensate) of gravitons at criticality  \cite{Dvali:2012en, Dvali:2013eja}.

   Within this framework,  the memory modes are the angular momentum modes of quantum fields which are made gapless by the black hole.  The emergence of these gapless modes can be understood as a result of criticality 
   of the graviton coherent state.
   The gapless modes can be described as the Bogoliubov/Goldstone modes of this state \cite{Dvali:2012en, Dvali:2015ywa,  Dvali:2015wca,  Averin:2016hhm}.      
   Such zero frequency modes are deposited by all existing particle species in the theory. 
   
In order to understand in QFT language the 
emergence  of zero-frequency modes  with very short-wavelengths, some important factors must be taken into account.  The fist point is that any black hole, regardless of its mass,   
breaks the Poincare symmetry spontaneously 
at the Planck scale. 
 
 \subsubsection{Spontaneous breaking of Poincare symmetry by a black hole}
     
    The Einstein gravity, viewed as field theory, is a theory of a massless spin-$2$  field,  
$\hat{h}_{\mu\nu}(x)$.  Choosing an asymptotically 
flat Minkowski space as the gravitational vacuum
(fully sufficient for our purposes), 
the quanta of the field $\hat{h}_{\mu\nu}(x)$
describe gravitons.  These particles obey 
an ordinary  Poincare-invariant dispersion 
relation between frequency and 
momentum,   $\omega_{\vec{p}} = |\vec{p}|$.

 The deviation of the classical metric from the flat one, 
$g_{\mu\nu}(x) = \eta_{\mu\nu} + \delta g_{\mu\nu}(x)$
is understood  as an expectation value 
of the quantum field over the corresponding quantum sate
of gravitons,
\begin{equation} \label{graviton} 
\delta g_{\mu\nu}(x) =  \frac{1}{M_P} \langle  \hat{h}_{\mu\nu}(x) \rangle \,. 
\end{equation} 
 In particular, the states that are well-described classically are the coherent states of high occupation number 
 $N$.  Correspondingly,  the basic point  of the  quantum
 $N$-portrait proposal   \cite{Dvali:2011aa}
 is that a black hole, at least at the 
length-scales of its horizon, is describable as 
 a coherent state of gravitons of wavelengths 
 $\sim R$ and the mean occupation number
 $N \sim S$.   In our terminology, these constituent 
 coherent gravitons are the master modes.

 Now, a black hole of mass $M$, placed in an asymptotically-flat Minkowski space, breaks the Poincare invariance spontaneously.   
 This feature is not exclusive to black holes.  Any localized macroscopic object (e.g., a soliton)  also breaks a part of the  Poincare symmetry spontaneously. 
 In particular,  the translations and the Lorentz boosts are 
 always broken. 
  
 However, what is special about a black hole is that the scale of breaking is given by $M_P$.  Indeed, the order 
 parameter  that determines the strength of 
 the spontaneous breaking  is the expectation value  of the canonically-normalized graviton field.   Far away from 
 the black hole, $r \rightarrow \infty$,  the expectation value 
 diminishes and departure from the Minkowski metric becomes less and less significant.   For example, at 
 distance $r$, the Newtonian component 
 drops as $\langle  \hat{h}_{00} \rangle \sim  
 M/(M_P r)$.  However, near the horizon, $r \sim M/(M_P^2)$,   this expectation value becomes of order $M_P$. 
 
  It is exclusively the property of a black hole that the 
 graviton expectation value reaches the Planck mass near the horizon.  For all other gravitating objects, 
  $ \langle  \hat{h}_{\mu\nu}(x) \rangle \ll M_P$ everywhere.   
 In particular, this is true for all astrophysical objects, 
 such as stars or galaxies. 
 
  The maximal breaking of Poincare symmetry by the black hole is what enables the emergence of
  modes with unusual dispersion relations that combine  
  the short wavelengths with zero frequencies.  
  In a Poincare-invariant vacuum, a mode 
  of wavelength $\sim 1/M_P$ would cost the energy gap
  $\sim M_P$.  By breaking the Poincare symmetry 
  at the Planck scale, the black hole creates an environment 
  in which the modes of arbitrarily short  wavelengths can be 
  gapless.

 \subsubsection{Diversity of memory modes and relation with species scale}

   The majority of modes created by the mechanism of the assisted gaplessness in $d$-dimensions come from the highest 
   angular momenta  $\sim \Lambda$, with their 
   number scaling as the area of the $d-2$-sphere 
  $\sim (R\Lambda)^{d-2}$ (see, \cite{Dvali:2017nis, Dvali:2018xpy}).   Since gravity couples to all the fields universally,  this counting applies to each QFT degree of freedom.

   Correspondingly, in a theory 
  with $N_{sp}$ species of quantum fields the total number of black hole memory modes can be estimated as,
  \begin{equation}  \label{NspBH}
    N_{\rm M}  \sim   N_{sp} \, (R\Lambda)^{d-2}  \,.  
  \end{equation}  
   The above expression may create a false 
   impression that the entropy of a fixed radius black hole  
  can be arbitrarily large, depending on the number of the QFT species in the theory.     
   This however is not the case, since 
    the cutoff $\Lambda$ depends on the number of 
  species non-trivially \cite{Dvali:2007hz, Dvali:2007wp}. 
    
    In order to see this,  we first carefully define the physical meaning  of $\Lambda$.   This  quantity marks  the scale above which gravity leaves the weak-coupling Einsteinian regime.  In pure Einstein gravity, with graviton the only 
   low-energy degree of freedom, the cutoff is given by the
   Plack scale, $\Lambda \sim M_P$. 
   
   However,  the black hole physics implies that in a theory in $d$ space-time dimensions  with $N_{sp}$ species of  $d$-dimensional quantum fields, the cutoff is lowered to the so-called ``species scale" \cite{Dvali:2007hz, Dvali:2007wp, Dvali:2008fd, Dvali:2008ec, Dvali:2008jb, Dvali:2009ks, Dvali:2010vm, Dvali:2012uq, Dvali:2021bsy},  
 \begin{equation} \label{Msp}
   \Lambda \, = \, \frac{M_P}{N_{sp}^{\frac{1}{d-2}}} \,. 
 \end{equation}    
  Of course, in the above expression $M_P$ has to be understood as the $d$-dimensional Planck mass.
  \footnote{Notice that indications for lowering of the cutoff 
  by number of species exist already in perturbation theory 
\cite{Dvali:2001gx, Veneziano:2001ah}. 
However, black hole argument is fully non-perturbative
and is insensitive to breakdown of loop expansion 
or resummation. }

   Substituting the relation (\ref{Msp}) into (\ref{NspBH}),  we obtain the following total number of the distinct gapless memory modes of a black hole, 
     \begin{equation}  \label{MspBH1}
   N_{\rm M} \,  \sim \, (RM_P)^{d-2} \,.  
  \end{equation} 
 This number  matches the  Bekenstein-Hawking entropy of a $d$-dimensional  black hole (\ref{BekH}), which is independent of the number of species. 
 Notice that the expression (\ref{MspBH1}) is independent of the details of UV-completion of Einstein gravity above the scale $\Lambda$, provided the 
 black hole size satisfies $R \gg \Lambda^{-1}$. 
 \footnote{ 
  In this respect it is worth mentioning that  
 an explicit realization of the formula 
 (\ref{MspBH1}) is provided by string theory,   where the role of the cutoff 
 $\Lambda$ is played by the string scale.
 In string theory the relation (\ref{NspBH}) translates 
 as the following bound on species in terms of the string coupling \cite{Dvali:2010vm}, 
 \begin{equation} \label{stringbound}  
  N_{sp} \leqslant \frac{1}{g_s^2} \,.
 \end{equation}
  This formula acquires a deep physical meaning 
  on $D$-brane backgrounds, where the species 
 originate from the Chan-Paton factors of the open strings. 
 Then one can see explicitly that the saturation of the bound  (\ref{stringbound})  makes the curvature radius 
 of order the string length \cite{Dvali:2010vm}. 
   In this limit, the relation (\ref{stringbound}) gets translated 
 into (\ref{MspBH1}). 
 
 Interestingly, the Chan-Paton species 
 play the role of the memory modes that can stabilize the 
 stack of $D$-branes via the memory burden effect.  
  This has been discussed in 
 \cite{Dvali:1999tq} in the context of the brane inflation
 \cite{Dvali:1998pa}.   More recently, the stabilization of $D-\bar{D}$-systems   by the  memory burden effect from the open string zero modes has also been studied in  \cite{Dvali:2024dzf}. There it was shown that  $N_{\rm M}$ number matches the Gibbons-Hawking entropy of the would-be de Sitter 
 state,  when the number of Chan-Paton species 
 saturates the bound (\ref{stringbound}).  Simultaneously, the curvature 
 radius $R$ approaches the string length and the 
 $D$-brane state gets stabilized by the open string memory burden.  This can be viewed as a string theoretic derivation 
 of the Gibbons-Hawking entropy \cite{Dvali:2024dzf}, 
 which complements the Strominger-Vafa derivation 
 of the Bekenstein-Hawking entropy of an extremal black hole \cite{Strominger:1996sh}.   
 Some other string-theoretic implications of the species scale can be found, e.g., in \cite{Dvali:2009ks, Dvali:2010vm, Dvali:2021bsy, vandeHeisteeg:2022btw, vandeHeisteeg:2023ubh, vandeHeisteeg:2023dlw, Cribiori:2022nke}
  }

    The recovery of  $N_{sp}$-independence of the black hole entropy represents an important consistency check of our microscopic framework.
 The black hole entropy comes out to be independent of the number of species because 
 the factor  $N_{sp}$  in (\ref{NspBH}) is exactly compensated by the $N_{sp}$-dependence of  the cutoff (\ref{Msp}).  
  \footnote{The $N_{sp}$-dependence of the cutoff (\ref{Msp})  also shows the independence of the entanglement entropy from the number of species \cite{Dvali:2008jb}. } 
       
    Thus, the diversity of the black hole memory modes, 
 $N_{\rm M}$, is independent of $N_{sp}$.   However, other features, such as the decay rate and the time-scale of the memory burden effect, do depend on $N_{sp}$.

 \subsubsection{Level-speading of the  memory modes} 
 
   Within our framework, the assisted gaplessness
   of the black hole memory modes can be summarized as follows.     
   In the asymptotic vacuum, due to Poincare symmetry,
 a mode of momentum $\Lambda$,  has energy $ \sim \Lambda$.   However,  in the vicinity of the black hole the Poincare symmetry 
  is spontaneously broken at the scale $M_P$.
  Correspondingly, the dispersion relation is modified and the memory modes become gapless while maintaining the large (angular) momenta $\sim \Lambda$. 
  That is, every gapless memory mode in the black hole spectrum 
 has an asymptotic counterpart of much higher energy.  
  
   We wish to make  a clarifying remark about the 
   gaplessness of the memory modes. 
   The point is that in quantum theory any localized object 
  represent a wave-packet which breaks the 
  translational symmetry of the Hamiltonian.  If the object is macroscopic, 
  meaning that the number of constituents is large, $N \gg 1$, the notion of spontaneous breaking 
  of the translational symmetry becomes well-defined.  
 That is, the states obtained by relative translations of the object can be viewed as the degenerate ``vacua" around 
 which the excitation modes can be
 quantized.  
 
  However, at finite $N$, this notion is only approximate, 
 since the shifted wave-packets have non-zero overlaps.  
  Correspondingly,  the Hilbert spaces formed by the excitations 
  quantized at relatively shifted locations are not exactly orthogonal. 
   Instead, the true Poincare-invariant ground state
   corresponds to the infinite superposition of the object at 
   all possible locations. 
   
    An isolated localized object, is not an eigenstate of the 
    Hamiltonian and shall evolve towards such a superposition. In other words, at finite $N$ (finite mass), 
     the quantum wave-packet shall start to spread. 
     
    For an object saturating the entropy bounds 
    (\ref{Area}), (\ref{Alpha}) and (\ref{Number}), the spread-out time is, 
    \begin{equation}  \label{spread}
      t_{\rm spr} \, \sim  \, N R  \, \sim \, S R \, .   
  \end{equation} 
  Of course, the above equally applies to a black hole. 
   The effect of this spread is that it introduces 
 a fundamental spread of the energy-levels given by
 \cite{Dvali:2021jto}, 
   \begin{equation}  \label{Lspread}
      \Delta {\mathcal E} \sim    
     \frac{1}{t_{\rm spr}}  \sim \frac{1}{N R}  \sim \frac{1}{S R}\, .   
  \end{equation} 
Correspondingly,  the lowest possible energy gap 
for any localized degree of freedom is given by the above expression.  This concerns also the black hole memory modes.  

 The expressions (\ref{spread}) and (\ref{Lspread}) 
 are of fundamental importance from the number of perspectives.  
 
 In particular, notice that all possible microstates are no longer exactly degenerate but rather are crowded 
within the energy gap $\Delta E \sim 1/R$.  This gap is of order 
the typical energy of a single Hawking quantum. 
 Correspondingly,  (\ref{Lspread}) also matches the spread 
 of the black hole  energy levels due to the black hole decay.  

 The expression (\ref{spread})  is also indicative from the point of view of the Page time \cite{Page:1993wv}, as it coincides with the latter.  
 This coincidence has been 
 explained previously (see, e.g., \cite{Dvali:2021jto}) and it has a deep physical meaning. 
 The expression (\ref{spread}) sets the upper bound 
 on the  time-scale over which the memory modes must start  to evolve. Correspondingly,  it tells us that in the absence of all other effects,  the black hole information would  start coming out after this time-scale.  
  This  supports the suggestion by Page but from a very different perspective. 

  With all the above said, for the purposes of our discussion, the gap
 (\ref{Lspread}) is  negligible, 
 since all the time-scales of our interest, such as 
 the swift memory burden time,  are much shorter than (\ref{spread}).
 We shall therefore continue to refer to the black hole memory modes as being ``gapless".  
 
  \subsubsection{Mapping to prototype Hamiltonians}

  While the information  stored in excitation 
  of a given $\Lambda$-momentum mode outside of a black hole costs energy $\Lambda$, the same information, stored 
   in a black hole, costs the (red-shifted) energy gap
  (\ref{Lspread}).  This difference creates the energy barrier that makes a fast  extraction  of the information from the black hole memory modes extremely difficult 
 \cite{Dvali:2018xpy, Dvali:2020wft}. 
   
   Thus, the same energy barrier,  between 
  gapless memory modes  and their asymptotic counterparts, that promotes the black hole into a device of 
   very efficient information storage, at the same 
 time, resists against the retrieval of the information. 
 
 As already pointed out in  \cite{Dvali:2018xpy, Dvali:2018ytn, Dvali:2020wft}, this feature is not specific to 
 black holes and gravity.   Rather, it is intrinsic to 
 the mechanism of the assisted gaplessness. It is therefore fully shared by the universality  class of objects that exhibit an  energy-efficient information storage capacity.   
 This universality allows us to understand the most important aspects of the memory burden phenomenon  using the prototype Hamiltonian (\ref{Hint})  with the proper mapping on the black hole  parameters.  
 
 This mapping goes as follows.  
 The role of the master mode is played by a soft coherent mode  $\hat{\alpha}$ of frequency $m_{\alpha} \sim 1/R$, which impersonates a graviton mode forming a classical near-horizon field of a black hole. This mode is occupied macroscopically to a critical number $N$. 
 This renders the memory modes gapless. 
  As already explained, the memory modes of a black hole, 
  which shall be represented by $\hat{a}$-modes,
  can be labeled by the spherical harmonics. 
 The effective Hamiltonians describing the assisted  
 gaplessness of the angular harmonics
  were given in \cite{Dvali:2017nis}, \cite{Dvali:2018xpy}.  It was also shown there that the spherical harmonic nature of 
 the memory modes immediately explains the area-law
 of the entropy. This is because the number of  gapless angular momentum modes scales as the area (\ref{MspBH1}).   
 The detailed construction, which can be found in the above references, shall not be repeated here.  
  For our purposes it suffices  to use the end-result of these 
  constructions  which is captured by the Hamiltonian of the type (\ref{Hint}).

\subsection{Swift memory burden}

 In order to describe the swift memory burden response 
 during a merger, we shall  analyse the two  interacting systems of the efficient information storage, one for 
 each black hole. 
The corresponding master and memory modes will be
dented by  $\hat{n}_{r} \equiv  
\hat{\alpha}_r^{\dagger}\hat{\alpha}_r$ and 
 $\hat{n}_{i}^{(r)} \equiv  
\hat{a}_{j(r)}^{\dagger}\hat{a}_{j(r)}$, respectively. 
The index $r=1,2$ labels the black holes, whereas 
$j=1,2,....M_r$ the corresponding memory modes. 
The analogous degrees of freedom of a final black hole 
will be denoted by similar symbols with index $r=3$. 
 The Hamiltonians describing the assisted gaplessness are
 \begin{equation} \label{Hr}
  \hat{H}_r = m_{r}  \hat{n}_r \, + \,    
  \left (1 -  \frac{
 \hat{n}_r}{N_r} \right )^p  \sum_j m_j^{(r)}  
  \hat{n}_j^{(r)} \,. 
 \end{equation}   
 In addition, we include 
 a mode  $\hat{\beta}$, with the number operator $\hat{n}_{\beta} \equiv  \hat{\beta}^{\dagger} \hat{\beta}$,
 describing the outgoing radiation  of frequency 
 $\omega_{\beta}$, with a free Hamiltonian 
 \begin{equation} \label{Hrad}
   \hat{H}_{rad} \, =  \, \omega_{\beta} \hat{n}_{\beta} \,.
     \end{equation} 
  The total Hamiltonian consists of five parts 
  \begin{equation}
    \label{Htotal} 
  \hat{H} \, =\, \sum_{r=1}^{3} \hat{H}_r \, +\, \hat{H}_{ rad}   \,
  + \, \hat{H}_{int}  \,,    
  \end{equation} 
 where $\hat{H}_{int}$ consists of all possible 
 interactions among various modes.  
 
 Of course, we must understand that in real black hole mergers  there exists 
 the whole tower of modes of each type.  In each
 event the system activates the ones that are most relevant for given set of parameters and the initial conditions. 
 Moreover, the activated modes change in time.  
    We therefore focus on such modes, ignoring the others. 
    
   Now, regarding the choice of the interaction 
   Hamiltonian $H_{int}$,  our task  is substantially 
   simplified by our target. Since our goal is to prove that 
   dynamics with and without memory burden are very different,  we can restrict to a simplest choice 
   of $H_{int}$  that makes this clear.  We therefore 
   take, 
                  \begin{equation} \label{HextraN} 
\hat{H}_{int} \, = \,  
 \tilde{m}_{\beta} \, \hat{\beta}^{\dagger} \hat{\alpha}_1 +  
 \tilde{m}\, \hat{\alpha}_3^{\dagger} \hat{\alpha}_2  
 \, + \, {h.c.} \,,    
 \end{equation} 
 where $\tilde{m}_{\beta}\,,  \tilde{m}$  are the mixing parameters. 
  We assume that the initial occupation numbers of the two merging master modes are critical $n_1 = N_1$, and  $n_2= N_2$.
 and initial occupation numbers in sector $r=3$ as well as in the radiation field are zero. 
 
   We also define the memory burden parameters for 
   the two black holes according to  (\ref{MuM}), 
   \begin{equation} \label{Mu12}
   \mu_{r} \equiv \frac{N_{r}m_r}{p E_{pr}}\,,~~~\, r =1,2. 
    \end{equation} 
 Now, let us compare the two evolutions with and without the memory patterns.  We first set the memory patterns in both initial black holes to be empty, $E_{p1} =  E_{p2} = 0$, 
 or equivalently, $\mu_1=\mu_2 =\infty$.  
Correspondingly, the  memory burden is never experienced  and the evolution proceeds in the following way \footnote{Of course, this is an idealized situation, since even if initially 
the black holes carry zero information loads,  some information loads will be picket up during the merger process. }.
  
   First, notice that without  any loss of generality we can assume  $m_1 \simeq m_3$  and $m_2\simeq \omega_{\beta}$. This is because in the tower of modes the system will automatically ``fish out" 
 the  $\alpha_{3}$ and $\beta$ modes with frequencies  that 
 are in resonance with the initial frequencies of
 the corresponding master modes, $m_1$ and  $m_2$. 
   At  each moment of time, the transition to the modes with higher level-splitting 
 is suppressed and can be ignored. 
  The evolution then is exactly solvable and shows  that 
  the occupation numbers of a new master mode 
  $n_{3}$ and  the radiation mode $n_{\beta}$ evolve in time similarly to (\ref{Nsin}),     
     \begin{equation} \label{Nsin1} 
     \frac{n_{\beta}(t)}{N_1}\,= \,\sin^2\left (\tilde{m}_{\beta}
t \right ) \,,~~~
\frac{n_{3}(t)}{N_2}\, = \,   \sin^2\left (\tilde{m}
t \right )\,. 
 \end{equation}
 That is, the initial master modes get fully converted into the new master mode and the radiation mode 
 over the time-scales  $t_1 \simeq \pi/(2\tilde{m}_{\beta})$ and 
 $t_2 \simeq \pi/(2\tilde{m})$ respectively. 
 
 Let us now consider the evolution with non-zero memory patterns $E_{p1}, E_{p2}$.  Since the behaviours in the two cases are similar, let us first focus on the evolution 
 of the radiation mode. 
   Again, we start with the critical occupation number 
   of the master mode $N_1$. And we assume 
   $p >1$, which guarantees that at the start of the time evolution the memory burden is absent, i.e., is not instantaneous.  
    
  The radiation mode gets populated as given 
  in (\ref{Nsin1}). 
  This evolution however continues 
  until the number of the radiation modes  $n_{\beta}$ (which is equal to the  number of the depleted master modes
  $\Delta n_1$) reaches the first threshold value, 
             \begin{equation} \label{critD1beta} 
 \frac{n_{\beta}}{N_1} \, \simeq 
\left ( \frac{N_1 |\tilde{m}_{\beta}|}{pE_{p1}} \right )^{\frac{1}{p-1}} \, = \,    \left ( \frac{|\tilde{m}_{\beta}|}{m_{1}}\mu_1 \right )^{\frac{1}{p-1}} \,.   
 \end{equation}
   This takes place after the time, 
              \begin{equation} \label{MBtime12} 
t_{\beta}\, \simeq 
\frac{1}{|\tilde{m}_{\beta}|}\left ( \frac{|\tilde{m}_{\beta}|}{m_1} \mu_1 \right )^{\frac{1}{2(p-1)}} \,.       
 \end{equation}
  Beyond this point, the memory burden sets in 
  and the transition amplitude to the mode of frequency 
  $\omega_{\beta} = m_1$
  gets suppressed as 
              \begin{equation} \label{Ampbeta} 
{\mathcal A} \, \simeq 
\left ( \frac{|\tilde{m}_{\beta}|}{m_1}\mu_1 \right )^2
\left(\frac{N_1}{n_{\beta}} \right) ^{2p-2} \,.      
 \end{equation}
 That is, the memory burden takes the frequency 
 $\omega_{\beta} = m_1$ off-resonance.  
After this, the depleting master mode must find 
an new resonant partner among the radiation modes, 
 with new values of $\tilde{m}_{\beta}$ and 
 $\omega_{\beta}$ that satisfy the resonant condition with the shifted frequency of the burdened master mode. 
 
  In general, the number of quanta on average released  into the radiation modes of frequency
  $\omega_{\beta}$, can be parameterized in terms of 
  an angle $\theta_{\beta}$ defined as 
                 \begin{equation} \label{thetaB} 
 \frac{\Delta n_{\beta}}{N_1} \, \equiv \sin^2(\theta_{\beta})  \,,      
 \end{equation} 
 which satisfies the following distribution,  
    \begin{eqnarray}       
    \label{thetaBS} 
  \tan(2\theta_{\beta}) \, &=& \, \frac{2\tilde{m}_{\beta}}{M(\theta_{\beta})}\, , 
 \end{eqnarray}
  where,   
  \begin{eqnarray}       
    \label{Mtheta} 
 M(\theta_{\beta}) &\equiv& m_1 \left (1 - 
 \frac{1}{\mu_1} \sin^{2p-2}(\theta_{\beta}) \right )  - \omega_{\beta}  \,.
\end{eqnarray} 
 
 The first term in brackets represents the effective frequency 
of the burdened master mode. 
The equation tells us that the memory burden 
shifts the resonant frequency towards infrared. 

Basically, the transition into a radiation mode 
of a given frequency 
$\omega_{\beta}$  
 can be unsuppressed only if this 
frequency  is in resonance with the 
effective burdened frequency 
of the master mode $M(\theta_{\beta})$.

 However, notice that the perfect resonance, $M(\theta_{\beta}) =0$, 
  that would give maximal depletion, $n_{1} = n_{\beta} = 
  N/\sqrt{2}$, is not possible for a pattern  with   
  the above-critical memory load,  
      \begin{equation} \label{EPstar}  
 E_{p1} >  \frac{1}{p} m_1N_1  \,,    
\end{equation} 
which corresponds to  $\mu_1 < 1$.

   For such patters, the memory burden is swift 
   and the resonances are very narrow. 
   Correspondingly, the time evolution quickly puts the system out of resonance.   
  In other words, even if at some given moment of time, the condition  $M(\theta_{\beta}) = 0$ is satisfied, 
 the further depletion violates the condition very fast.

 In order to see this, we must solve the equations self-consistently.   Since  $M(\theta_{\beta})\geqslant 0$, 
 the equation  (\ref{Mtheta}) tells us that 
 $\sin^2(\theta_{\beta})$ is bounded from above by the 
 condition, 
                  \begin{equation} \label{thetaBound} 
 \sin^2(\theta_{\beta}) \leqslant \left ( \mu(1- \frac{\omega_{\beta}}{m_1} ) \right)^{\frac{1}{p-1}}\,.      
 \end{equation} 

  For a typical memory burden (\ref{EPstar}),  this quantity 
  is very small, and the equation 
  (\ref{thetaBS}) implies, 
     \begin{eqnarray}       
    \label{thetaSmall} 
   \sin^2(\theta_{\beta})&=& \frac{\tilde{m}_{\beta}^2}{M(\theta_{\beta})^2}\, \ll \, 1\,, 
 \end{eqnarray}
which confirms that the resonance is narrow and 
the number of radiated quanta is small. 
  Thus,  the memory burden is swift and it strongly affects the radiation dynamics. 
  
   The equation (\ref{thetaBound}) represents the 
   fraction of radiated quanta (intensity) in given frequency 
   range $\omega = \omega_{\beta}$.  Inserting the black hole parameters, 
    $E_P = M_P N_p,~N = S = (M_PR)^2,~ m_1=1/R$, we can rewrite this equation  in the form,  
 \begin{eqnarray} \label{Intesity} 
  I_{\omega} &\lesssim& 
  ((1- R\omega) \, \mu)^{\frac{1}{p-1}} \,,
\end{eqnarray}
 where the memory burden parameter  (\ref{MuM})
 for a black hole 
 of mass $M = m_{\alpha} N$  is,  
    \begin{equation}   \label{MuBH}
 \mu \, \sim \,  \frac{M}{E_P}  \,.
 \end{equation}

   We thus observe the following features: \\
   
    {\it 1)}   The suppression of the spectrum is controlled 
   by the memory burden parameter $\mu$.  \\

   {\it 2)}   The spectrum is moved 
   towards infrared. Of course, here we must take into account that, since for the lower frequency modes 
   the mixing terms, $\tilde{m}_{\beta}$, are smaller,   
   the intensity will be further suppressed due to   (\ref{thetaSmall}).  \\

 The above behaviour is universal for a transition of  a memory burdened mode into the modes of frequency $\omega_{\beta}$. Correspondingly, the second master 
 mode $\hat{\alpha}_2$ transiting into  a new master mode 
 shall experience a similar evolution.  
 This can be described by replacing the labels  $1\rightarrow 2$, 
 $\beta \rightarrow 3$ in  the above equations.

Of course, in reality the time evolution is more complicated 
 since the modes mix non-trivially.  However, the qualitative 
 features are so obvious that they allow us to  derive 
 a master formula indicating the fraction of the energy 
 budget that will get invested into  a new dynamics 
 due to the memory burden response. 
  This fraction is determined by the black hole memory burden  parameter  (\ref{MuBH}).

\section{Memory burden effect on spectrum of black hole excitations}  

 Let us estimate the general  
 spectrum of black hole perturbations  subjected to the memory burden effect. 
    In general, a quantized perturbation of gravitational field  can be written as   
    \begin{equation}  \label{BHpert}
    \hat{h}_{\mu\nu} (\vec{x},t) = \sum_{\alpha} \frac{1}{\sqrt{2m_{\alpha} V}}  \psi_{\alpha}(\vec{x}) {\rm e}^{im_{\alpha} t} \,  \hat{\alpha} \, \epsilon_{\mu\nu}^{(\alpha)} + {\rm h.c.}  \,,
    \end{equation}  
 where, $\hat{\alpha}$ are the relevant modes, 
 $\psi_{\alpha}(\vec{x})$ are the corresponding mode-functions, and $\epsilon_{\mu\nu}^{(\alpha)}$ the polarizatio tensors.  $V$ is the relevant volume of the system.   
 The classical properties of the background are encoded in the mode-functions  $\psi_{\alpha}(\vec{x})$, which also determine the dispersion relations $m_{\alpha}$. 
  For small perturbations, in general, it is convenient to choose the $\alpha$ labels according to the symmetries of the background.   For example, for linear perturbations 
  on top  of a stationary Kerr metric, the labels 
  $\alpha$ can refer to spherical harmonics $l,m$ and 
  overtones.  
 
   In order to avoid a potential confusion of our description with the effects currently discussed in the standard spectroscopy of black hole  perturbations (see, e.g., \cite{BHspectrum3}), some comments are in order.   
   
 First, regarding notations, although we can still use the spherical harmonics 
 as labels, our description is somewhat different as 
 we resolve the background itself as a coherent state 
 of the constituent gravitons in the spirit of \cite{Dvali:2011aa}. 
  Notice that one does not have to rely on this 
  microscopic picture,  as  the memory 
  burden effect is independent of it;  we merely use 
  the coherent state description of the classical gravitational 
  field as the most convenient 
  setup for explaining the essence of the effect. 
  
  Secondly, we are after the effects that are higher order 
  in quantum correlations.  It is customary to  assume 
  that such effect are unimportant for the classical dynamics
  of the mergers.  Our goal is to challenge this view and bring the awareness that   
for a perturbed black hole, the effect of these correlators 
 can be macroscopic as it is amplified by the memory burden  parameter $\mu^{-1}$.

   Next, we restrict our attention to the master modes.  Correspondingly, $\hat{\alpha}$-s will stand for  
  the graviton degrees of freedom that are most relevant 
  for black hole's near-horizon classical field viewed
  as a coherent state. For simplicity, 
   we shall take the volume to be 
 given by the black hole scale $V \sim R^3$.
 Since, in general, we allow for dissipation, 
 the frequencies $m_{\alpha}$ can have imaginary parts. 
 
It is important to clearly stress that in our estimate the  
existence of a graviton coherent state describing  a black hole's  near-horizon physics is an input assumption. 
It is also assumed that a set of memory modes $\hat{a}_j$
 is becoming gapless in such a state.  These are obvious assumptions that are necessary for describing a black hole as a legitimate state in gravity as well as for accounting 
 its entropy.  As already discussed, the memory 
 modes can be labeled by the spherical harmonics and 
  additional species labels.
 
  In this setup, we wish to explore the back-reaction from 
 the information pattern carried by the memory modes on the coherent perturbations of the master modes. 
 
 In order to directly connect with the previous analysis performed with the prototype Hamiltonians, it is most convenient  to  view the black hole's near-horizon field 
 as an expectation value over a coherent state of gravitons constructed on top of the 
 asymptotic Minkowski vacuum:  
   \begin{equation} \label{coherentH}
   \ket{C} = {\rm e}^{\sum_{\alpha} c_{\alpha} \hat{\alpha}^{\dagger} - c_{\alpha}^* \hat{\alpha}} \ket{0} \,,
  \end{equation} 
   where the summation  goes over all the involved master modes and the coherent state parameters, $c_{\alpha}$, are complex numbers that set the mean occupation numbers 
   of the master modes via  $|c_{\alpha}|^2 = n_{\alpha}$. 
 We shall only use the most basic features of this description. 
  Explicit constructions of graviton coherent states 
  including the BRST-invariant ones can be found in 
 \cite{Dvali:2017eba} and in  
  \cite{Berezhiani:2021zst, Berezhiani:2024boz, Berezhiani:2024pub} respectively. 
 
 Now, the key point is that the state 
 $(\ref{coherentH})$ accounts only for  the coherent 
 part of the black hole's ground-state. In particular, 
 it does not capture the physics of the memory modes, which in the ground-state has a very little effect on a black hole.   However, 
 one can no longer ignore the memory modes when the 
 black hole is perturbed.  Via higher order correlators, 
 they exert the back reaction  on the coherent fluctuations of the master modes.  
 
 In our proof of concept estimate, we shall only consider the effects that  are summed up in an uniform memory burden parameter $\mu$.  In more refined analysis, 
 one can spectrally decompose the effect
 according to the harmonics of the contributing 
 memory modes.

    Obviously,  the constituent master modes 
   $\hat{\alpha}$  are off-shell relative to their asymptotic counterparts. However, unlike the memory modes, 
    in the black hole ground state the frequencies of the  
    master modes $m_{\alpha}$ are comparable to their inverse wavelengths.   
   That is,  in a state of unperturbed black hole, viewed as a coherent state of the master 
   modes, 
    the bulk of the near-horizon classical dynamics 
 is taken-up  by the modes 
    with frequencies $m_{\alpha} \sim 1/R$. Their  
    total occupation number is $\sum n_{\alpha} = N \sim S$. 
    These modes are the ones mainly responsible for the  effect of the assisted gaplessness.  
     The contributions from higher and lower frequency 
     modes are sub-leading.          
     Correspondingly, these are the modes whose classical dynamics is most affected by the information pattern carried 
     by the memory modes when the black hole is perturbed.     
   
      Therefore, for our estimates we adopt the following simplified picture, which however captures the key aspects of physics. 
  The black hole coherent state consists of master modes 
  of characteristic wave-lengths $\sim R$ and  frequencies $m_{\alpha} \sim 1/R$.  These are occupied to 
  critical numbers $n_{\alpha}$  which render the memory modes gapless. 
  
   We study the coherent perturbations of the graviton field 
   around this critical state. 
    In coherent state description of classical gravitational field, these perturbations are mapped to perturbations  in the occupation numbers of the master modes $\Delta n_{\alpha}$.  Of course, these are linked with the perturbations of  the coherent state parameters $c_{\alpha}$.   
    
    Therefore, the following expression for the amplitude 
     of perturbation of the corresponding $\alpha$-harmonic of the gravitational field, 
     \begin{equation}
   \delta h_{\alpha}^2 \equiv  
    \bra{n_{\alpha} + \Delta n_{\alpha}}  \hat{h} 
    \ket{n_{\alpha} + \Delta n_{\alpha}}^2  - 
    \bra{n_{\alpha}}  \hat{h} 
    \ket{n_{\alpha}}^2 \,,
    \end{equation}      
     satisfies, 
     \begin{equation}  \label{BHpert1}
    \delta h_{\alpha}^2 \sim \frac{\Delta n_{\alpha}}{R^2} 
    \frac{1}{m_{\alpha R}}  \,.
    \end{equation} 
      Dividing both sides by  $M_P^2$ and taking into account the relation  $(M_P R)^2 \sim S \sim N$, we 
   translate the above in the relation of the dimensionless metric perturbations, 
     \begin{equation}  \label{BHpert2}
    \delta g_{\alpha}^2\,  \sim \,  \, \frac{\Delta n_{\alpha}}{N}
    \frac{1}{m_{\alpha R}}  \,.
    \end{equation} 

  Now,  using our master formula (\ref{nphiA}), 
we find the critical amplitude of the perturbation above 
which the memory-burden effect is unavoidable, 
     \begin{equation}  \label{BHpert3}
    \delta g_{\alpha}^2 \sim 
    \frac{1}{m_{\alpha R}} \, \mu^{\frac{1}{p-1}}  \,.
    \end{equation} 
    
   The above expression 
  is applicable to a generic perturbation of the black hole's 
  classical field. In particular, it can be viewed as the memory-burden  constraint on the quazinormal modes.

  For perturbations with $m_{\alpha} \sim 1/R$, which are 
   most relevant during the mergers, we have, 
    \begin{equation}  \label{BHpertA}
    \delta \hat{g}_{\alpha}^2 \, \sim  \, \mu^{\frac{1}{p-1}}  \,.
    \end{equation} 
    The perturbations exceeding the above  critical value must be subjected to a full memory burden effect. 
  Notice that during the mergers, the amplitude of perturbations is order-one. Correspondingly, the 
  effect of the memory burden on mergers of black holes with $\mu \lesssim1$, must be significant.  
       These effects have to be imprinted in the spectrum of gravitational waves at corresponding wavelengths.        
   Therefore the  swift  memory burden effect can be detected 
  via gravitational waves thought its imprints in the black hole spectroscopy \cite{BHspectrum1, BHspectrum2} (for a recent review and updates, see 
 \cite{BHspectrum3}).

 It is important not to confuse the corrections coming from the swift memory burden effect  with the ones coming 
 from classical non-linearities.  Such non-linearities 
 can be accounted by the one-point function of the graviton field.  In contrast, the memory burden effect comes from higher-point functions which are enhanced due to the macroscopic 
 memory burden parameter $\mu^{-1}$.  
 
  Because of the importance of the point, we summarize  
  it again in the following way.  The black hole state $\ket{BH}$ consists 
  of  the coherent part composed  by the master modes 
  (\ref{coherentH}) and the information pattern 
  of the memory modes. For simplicity, putting aside the possible entanglement among the memory modes, 
  we can write, 
  \begin{equation} \label{BHstate}
  \ket{BH} = \ket{C} \times \ket{n_{1}, n_{2}, ....} \,.
 \end{equation}   
     In the black hole ground state, physics is mainly 
     described by the classical gravitational field  which is 
     accounted by a one-point function of the graviton 
     over the coherent state of the master modes 
     $\ket{C}$.  The contribution from the information pattern 
     of the memory modes, comes from higher-point 
 functions and is $\sim 1/N \sim 1/S$, since the 
 memory modes are essentially gapless (\ref{Lspread}).
 
 If the black hole is not perturbed classically, the         
  higher-point correlators will grow slowly  due to the 
  back-reaction from the Hawking evaporation, which 
  per emission time is $\sim 1/S$ \cite{Dvali:2013eja, Dvali:2015aja}. 
  This back-reaction gradually increases  the gaps of the memory modes and so does their contribution 
  into the higher-point functions.  Eventually, after a certain macroscopic time, the  higher point functions will reach the level of competition with the one-point ones and the   
  memory burden will set in fully \cite{Dvali:2018xpy, Dvali:2020wft}.    
   
   Notice that  the phenomenon of a departure from classicality, the 
   so-called  ``quantum breaking" \cite{Dvali:2013vxa}, can take place  in  time evolutions of generic coherent states
  \cite{Dvali:2013vxa, Dvali:2013eja, Dvali:2015wca, Dvali:2016zqx, Dvali:2017eba, axions, MarcoSeb}. 
   In particular,  the analysis  for such departures via higher correlators in scalar theories has been given in \cite{Lasha1, Lasha2, Lasha3, Lasha4}.  Of course, in a generic system, the effect of quantum breaking is not necessarily linked with the memory burden effect, which is our main focus. 

 However, the converse is always true: the memory burden effect inevitably influences the classical evolution 
 of the system.  In particular, this influence is swift if the 
 black hole is perturbed classically.  
 Such a perturbation increases the gap functions of the memory modes, and correspondingly, their contributions 
 into the higher-point functions. 
 If the information load is significant, this contribution is 
 enhanced macroscopically. 
 In other words, in case of the swift memory burden, 
 the main source of the quantum-breaking is the 
 contribution from the memory modes rather than the 
 break-down of coherence in the time-evolution of the master mode.

 \section{Estimating the memory burden parameter}
 
  As we have seen, the strength of the imprints of the swift memory burden effect activated during the merger is controlled 
  by the black hole memory burden parameter $\mu$ (\ref{MuBH}). 
     It is therefore important to have some estimates 
     of this parameter, especially for astrophysical 
     black holes 
     which are the sources of  observationally-accessible gravitational waves. 
     
 The memory load parameter tells us what fraction 
 of the  information-storage capacity is used by the 
 information-load  
 actually carried by a given black hole.  
  That is, it measures the fraction of the memory modes 
  occupied by the memory pattern weighted  by their momenta.  
  
  The memory burden parameter depends on the information carried by the collapsing source.
   The absolute lower bound 
   of the memory burden parameter, expressed via the black hole entropy,  is, 
   \begin{equation} \label{minMu} 
    \mu \gtrsim \frac{1}{\sqrt{S}}  \,.
   \end{equation} 
  This limit is reached in the extreme case in which the 
  information carried by the collapsing source takes up the entire storage capacity of a black hole. 
  That is, the information is encoded into the memory modes 
  of maximal diversity which have the shortest possible wavelengths, $\sim 1/M_P$. The corresponding memory pattern satisfies $E_p \sim  S M_P$, which from 
  (\ref{MuBH}) gives (\ref{minMu}). 
  
  Theoretically, if the black hole information pattern is empty, 
  $\mu$ can be infinite. In practise this can never happen 
  since the collapsing source always carries non-zero information.  
  The  observationally-interesting  values are $\mu \lesssim 1$. For $\mu \ll 1$ the swift memory burden effect is substantial.

    We shall now derive some simple master formulas for 
   $\mu$.    
      In order to do this, let us consider a region of radius $R_0$   
     filled with matter that collapses into a black hole. Unless the region is exceptionally-symmetric,  a fraction of its energy shall be radiated 
   away during the collapse. 
     We shall assume that this fraction is 
   of the same order or less than the initial energy 
   of the region.   In other words, we assume that 
   the mass of the resulting black hole, $M$, is of the same order  as the energy of the collapsing matter. 
   
    During the collapse, the     
   radiation will carry away a part of the information
   encoded in the initial state.  The remaining 
   fraction will be encoded into the black hole, predominantly 
   in form of the memory pattern. 
    Again, for simplicity we assume that the 
    fraction of information inherited by the black hole 
    is significant.

   Let the initial number of the excited degrees of freedom be 
   $N_p$ and their characteristic frequencies be 
   $\omega_0$. 
   The mass of the collapsing region 
   which  is (approximately) equal to a mass of a black hole 
   then is $M \sim N_p \omega_0$.   Let us assume that 
   after the collapse the memory pattern gets rewritten 
   in black hole memory modes of  angular momenta 
   $m_j$ given by  some characteristic scale, $m_j \sim m_{\rm M}$.

   Then, let us distinguish two cases.
    First, we assume that the collapsing matter 
   has maximal diversity. That is, all exited 
   degrees of freedom are in distinct one-particle states
   distinguished say by momenta, spin and other quantum numbers.  
      Under this assumption, the diversity of  quanta 
      is equal to their total occupation number 
      $N_p$.

   In such a case, the black hole memory pattern, to which the information gets encoded, satisfies 
    $E_p \sim N_p  m_{\rm M}$. 
   From  (\ref{MuBH}) we then get the following simple formula 
  for the black hole memory burden parameter,   
     \begin{equation} \label{MBdiver}
   \mu \, \sim \, \frac{\omega_0}{m_{\rm M}}\,.
   \end{equation} 
   
 It is easy to argue that the above memory burden parameter
 must be less than one. First, by assumption of maximal initial diversity, the diversity of modes  with absolute value of momentum $|\vec{p}| = \omega_0$ scales as
 \begin{equation} \label{NPinitial}
 N_p \, \sim  \, N_{sp} (\omega_0 R_0)^3 \,, 
 \end{equation}
 where $N_{sp}$ accounts for additional 
  degeneracy of species with respect to the other quantum labels.

  The information carried by this diversity must be 
  accommodated  as the memory pattern of the black hole 
  memory modes  of angular-momenta $\sim m_{\rm M}$. 
 The diversity of such memory modes scales as 
 $N_p \sim (m_{\rm M} R)^2$.   
 This gives the following inequality,
  \begin{equation} \label{2NPs}
 (m_{\rm M} R)^2 \,   \gtrsim\,  N_{sp} (\omega_0 R_0)^3 \,.
\end{equation}  
 Correspondingly, the memory burden parameter 
 satisfies, 
        \begin{equation} \label{MBMBd}
   \mu \, \lesssim \,  \frac{1}{\sqrt{N_{sp}}}\frac{R}{R_0} 
 \frac{1}{\sqrt{\omega_0 R_0}}  \ll 1 \,,
   \end{equation} 
   Since, $R < R_0$,  $N_{sp} \geqslant 1$ and 
   $\omega_0 \gg 1/R_0$,  the memory burden parameter 
   is typically much smaller than one. 
      
 Also, notice that the above is just an upper bound obtained 
 without taking into account the dynamics of the 
 encoding mechanism.  This can further increase 
 $m_{\rm M}$, since the diversity of
 short wavelength memory modes is much higher and 
can increase the probability of encoding the initial information in such modes.  This  will further decrease $\mu$. 
   
  We thus conclude that for a black hole 
  formed via a collapsing matter of maximal diversity, the 
  memory burden parameter is small. Upon an external 
  disturbance, such black holes are 
  pre-disposed to a swift and strong memory burden effect.

   High diversity is generic for collapsing sources of a 
 relativistic matter  of some quantum fields of energy density 
  \begin{equation} \label{Cmatter} 
  \rho_{\rm in} \sim N_{sp} \omega_0^4 \,,
  \end{equation} 
  where $N_{sp}$ is the number of actualized QFT species. 
   The corresponding number of the activated
  degrees of freedom is $N_p \sim  N_{sp} (\omega_0 R_0)^3$ which can be expressed through the entire energy of the region, $M \sim N_{sp} \omega_0^4 R_0^3 $,  as  
  $N_p =  \frac{M}{\omega_0}$.
  In such a case the memory burden parameter is 
  given by (\ref{MBdiver}) and satisfies 
   (\ref{MBMBd}). 
   
      As an illustrative example,  for a few solar mass black hole formed by a collapse of a mildly relativistic matter 
  of nuclear density, the memory burden parameter  satisfies 
         \begin{equation} \label{SolarMu}
   \mu \, \lesssim \,  10^{-9}  \,,
   \end{equation} 
which is a significant memory burden.

 Let us also discuss PBH ~\cite{Zeldovich:1967lct, Hawking:1971ei,Carr:1974nx,Chapline:1975ojl,Carr:1975qj}. 
     An example of a black hole obtained by 
     a high-diversity source, is provided by a PBH formed 
     in a collapse of a thermal bath of temperature $T$. 
     The memory burden parameter satisfies 
     (\ref{MBMBd}) with the substitution $\omega_0 = T$.

  On the other hand, for a PBH obtained by a collapse of a  Hubble region,  the story is less certain.  The memory pattern of such black holes  must accommodate the information carried 
 by the de Sitter memory modes \cite{Dvali:2018ytn}
 that are  responsible for the
 Gibbons-Hawking entropy \cite{Gibbons:1977mu}.  This memory-load 
 depends on the initial conditions and can be close to maximal, implying  $E_P \sim S M_P$ \cite{Dvali:2018ytn}. 
 For such a black hole, the memory burden parameter is, 
      \begin{equation}   \label{MuXi}
 \mu = \frac{1}{\sqrt{S}}  \,,
 \end{equation}   
 and swift memory burden can be close to maximal. 
    
    The opposite case is when the collapsing source has a low 
      diversity.  This is the case if the energy of the source comes from high occupation numbers of identical quanta and there are no gapless memory modes present. 
      
       For example, the role of such a low diversity 
source can be played by an under-critical  Bose-Einstein condensate in a state of a very low micro-state degeneracy.
       If such a source collapses into a black hole,  the resulting memory burden parameter can be large and the effect of the burden insignificant. 
       
      However, it is important to stress that Bose-Einstein condensates in interacting theories can exhibit an extremely 
 high diversity due to the assisted gaplessness \cite{Dvali:2017nis, Dvali:2018xpy}. 
      Such condensates can even saturate the bounds 
      (\ref{Area}), (\ref{Alpha}) and (\ref{Number})
      on the microstate degeneracy. 
   
     A many-body example of such a critical condensate was introduced in  \cite{Dvali:2012en}  as a simple prototype model for a black hole graviton condensate of $N$-portrait.
  This example  shall be reviewed later in connection with possible laboratory studies of the memory burden effect. 
  
  Since, such a condensate itself represents  a system of high-efficiency of information storage, it can carry a significant memory pattern.  Correspondingly, if 
  an object composed out of such condensate collapses 
  into a black hole, the memory pattern will get  encoded into 
  the black hole memory modes of corresponding high diversity, resulting into a highly suppressed burden 
  parameter.  This will manifest itself in swift memory burden 
  effect.

  The standard matter sources responsible for the formation 
 of astrophysical black holes  carry sufficient diversity 
 for endowing  the black holes with a very significant 
 memory load.  Such black holes are expected to 
 exhibit the swift memory burden effect 
 during mergers and other perturbations. 
 
  The same applies to PBH obtained by the 
  collapse of a radiation bath.  
  This is especially true for PBHs that form 
  in a collapse of a Hubble region filled with radiation of 
  temperature $T$.
  Such PBHs are expected to carry a maximal 
  information load.  This is because of the following 
  reasons. First, the diversity of the thermal matter 
  within the Hubble patch is given by,
   \begin{equation} \label{NMHubble} 
  N_{\rm M} \sim N_{sp} (R_H T)^3 \,,
  \end{equation} 
 where $R_H \sim M_P/(T^2\sqrt{N_{sp}})$ is the 
 Hubble radius.   It is easy to see that this diversity 
 fully matches the entropy of a black hole obtained by the 
 collapse of the Hubble patch which has the radius 
 $\sim R_H$, 
  \begin{equation} \label{NMHubble} 
  N_{\rm M} \sim  S \sim (R_H M_P)^2 \,.
  \end{equation} 
 Correspondingly,  we can  conclude that the memory burden parameter of PBH formed in a collapse 
 of a radiation-dominated Hubble patch is (\ref{MuXi}) and such PBH must carry a maximal memory burden.  
  Notice,  this memory load also matches the inherited information capacity of the de Sitter space given by the Gibbons-Hawking entropy \cite{Gibbons:1977mu}.

    In conclusion of this chapter,  the memory burden parameter of a black hole 
     is determined by the information load carried 
   by the  collapsing source, which can be expressed 
   in terms of the diversity of the excited modes. 
  Simple estimates indicate that standard 
  sources responsible for formation of astrophysical 
  black holes have sufficient diversity for 
  creating a significant memory load. 
  Correspondingly,  perturbations of such black holes 
  are expected to experience a substantial swift memory burden effect. 
  
  \section{Compatibility with black hole's classical properties}
 
   Classically,  black holes  satisfy the so-called 
   no-hair theorems \cite{NH1, NH2, NH3, NH4, NH5, NH6, NH7}. 
    They state that a static black hole can be fully characterized by its mass, its charge (electric and/or  magnetic) and the angular momentum. This set can be extended to the charges under the additional massless gauge fields, if such exist in the theory.
  These no-hair features are in  complete agreement with  the property that the  information stored in  a classical black hole is unreadable.  
  
  However, in quantum theory, the black hole acquires a 
  hair.  The microscopic origin of this hair was originally discussed in the $N$-portrait description of a black hole
   \cite{Dvali:2011aa}.  
    As already discussed, according to this picture, at  the initial stages of evaporation the hair (information) is stored in corrections of the strength  $\sim 1/N \sim 1/S$  \cite{Dvali:2012rt}. 
   This fully matches the general lower bound 
 (\ref{Gbound}) on the deviations from thermality in the spectrum of Hawking radiation  which is independent of particularities of the  microscopic theory \cite{Dvali:2015aja}.

   Of course, for a large black hole,  the value of $1/S$ 
   is tiny.  This may create a false impression that the effect 
   is unimportant. 
   First, as already discussed,  the $1/S$-effect are resolvable over the  time-scale (\ref{spread}). 
    Most importantly, the back-reaction effect is cumulative, and grows in time. One must keep in mind that 
 for larger $S$, the black hole correspondingly lives longer. 
 Even with a naive extrapolation of the semi-classical 
 result, the half-decay time scales as (\ref{Thalf}) which also matches (\ref{spread}).
  Correspondingly, the latest by the time (\ref{spread}),  an order-one  
impact from the back-reaction becomes unavoidable. 
 
   A particular manifestation of the black hole's  quantum hair is the memory burden effect \cite{Dvali:2018xpy, Dvali:2020wft}. This effect tells us that, in quantum theory, 
 a black hole  has a new macroscopic characteristics in form of  the memory burden parameter $\mu^{-1}$ which measures  its information load.
 
  The unusual thing about this parameter is that, while 
  it is quantum in origin, it leads to the macroscopic 
  effects.  This phenomenon represents a particular manifestation of what in \cite{Dvali:2012wq}
was called  the black hole's 
  ``macro-quantumness".  
     The macroscopic nature of the parameter $\mu$ is 
   apparent from the fact that for a black hole with 
   maximal memory load (\ref{minMu}), the product
   $M\mu \sim M_P$ diverges in the classical limit.  
      
   In the previous  studies,  the manifestations of the memory burden effect were discussed after a macroscopic time reached in the process of a gradual evaporation.   
  
   The swift manifestation of the memory burden effect, discussed in the present paper,  tells us that
   $\mu$  can have the immediate macroscopic 
  effects which can change the classical dynamics of a 
  perturbed black hole.

    The black hole's memory burden effect 
    is of course in no conflict with the classical no-hair 
    theorems.  These theorems 
   restrict  the parameters of static classical black holes. 
    The parameter $\mu$ is quantum in origin 
    and it manifests itself only for a 
    perturbed and thus time-dependent 
    black hole.

    It is interesting to ask how  the swift memory burden 
    effect could correct other  known classical properties
    of a black hole.  For example, Hawking's area theorem  \cite{Hawking:1971vc} and its extensions \cite{HArea1, HArea2, HArea3}  state that classically the horizon area can only increase. One can ask whether the swift memory burden effect could change this. Of course, since the 
 effect involves a macroscopic parameter $\mu$ that is quantum in origin, a more careful analysis is required. 
  However,  at least from the first glance, it appears that the memory burden 
  effect would only help in maintaining the growth of the horizon area, since such a growth increases the memory space. 
   However, the dynamics of the growth will of course be affected,  since the time evolution  cannot  be confined to the  null memory burden  surface.


    \section{ Swift Memory Burden Phenomenon
    in Solitons} 
     
  We shall now illustrate the swift memory burden 
  effect  in QFT solitons.  For conceteness,  we shall use the explicit example of a high  
 information capacity soliton constructed in  \cite{Dvali:2019jjw}.   A detailed review and analysis of the model can also be found in \cite{Alessandro}.  
 
    The soliton in question is a 't Hooft-Polyakov 
    monopole. It was shown in  \cite{Dvali:2019jjw} that this object can  be endowed by a maximal capacity of information storage, saturating the bounds (\ref{Area}), (\ref{Alpha})  and (\ref{Number}) on the microstate entropy.  
    
    We shall first repeat the steps of the construction 
    and later illustrate the swift memory-burden 
    effect.  The burden is activated  when the monopole is subjected to an external classical disturbance. 
    Such disturbance can come, for example, in form of a merger with an anti-monopole.    The macroscopic differences 
    between the behaviours of the system with and without 
   the memory burden shall become very transparent.

 Following \cite{Dvali:2019jjw}, let us consider a simple theory that contains a 't Hooft-Polyakov monopole \cite{Monopole}. 
 This is a theory with a gauge  $SO(3)$-symmetry 
 spontaneously broken (``Higgsed") down to
 its $U(1)$-subgroup by a non-zero vacuum expectation  value (VEV) of a $SO(3)$-triplet scalar field  $\Phi^a$, with 
   $a = 1,2,3$  the $SO(3)$-index.  The Lagrangian
   has the following form:  
     \begin{eqnarray}   \label{Lag1} 
  &&  L =  {1 \over 2} D_{\mu}\Phi^a D^{\mu}\Phi^a
  - {1 \over 4} F_{\mu\nu}^aF^{\mu\nu a}
     -  \\ \nonumber
   && - {h^2 \over 4} (\Phi^a\Phi^a - v^2)^2 \,,     
 \end{eqnarray}
  where  $D_{\mu}\Phi^a \equiv \partial_{\mu}\Phi^a
  + e \epsilon^{abc} A_{\mu}^b\Phi^c$ is the covariant derivative and 
 $F_{\mu\nu}^a \equiv \partial_{\mu}A_{\nu} ^a -\partial_{\nu}A_{\mu} ^a
  + e \epsilon^{abc} A_{\mu}^bA_{\nu}^c$ the field-strength. 
 The parameter $v$ has dimensionality of  mass, whereas the   parameters 
  $e$ and $h$ are dimensionless gauge and Higgs coupling constants respectively.

 In the topologically-trivial vacuum,  the Higgs VEV can be chosen as
 \begin{equation} \label{Higgsvacuum}
   \Phi^a  = \delta^{a3} v  \, .
  \end{equation} 
   This VEV Higgses the $SO(3)$ gauge group down to its Abelian  $U(1)$-subgroup of rotations in the $1-2$ plane.   
  The corresponding gauge boson $A_{\mu}^3$ remains massless, whereas the two other gauge bosons $A_{\mu}^{1,2}$ gain masses equal to $m_v = ev$.

The Higgs boson, a scalar degree of freedom that describes fluctuations of the absolute value of the VEV, gains the mass $m_h = hv$.

  In addition to the topologically-trivial vacuum, 
  there exist monopole solutions with a non-zero magnetic charge.    
   The 't Hooft-Polyakov monopole  of the unit charge is described by the  solution of the following form
 \cite{Monopole}
\begin{equation}\label{Mon}  
 \Phi^a = {x^a \over r} vH(r), ~~ A_{\mu}^a = {1 \over er} 
 \epsilon^{0a\mu\nu} {x_{\nu} \over r} F(r)\,,
 \end{equation} 
 where $r$ is the radial coordinate.  
The asymptotic values of the two functions are 
$H(0)=F(0)= 0,\, ~H(\infty) = F(\infty) = 1$. 
The monopole is invariant under the combined 
$SO(3)$-rotations in coordinate and internal spaces. \\

 Although the magnetic flux of the monopole extends to infinity, the monopole radius $R_{mon}$ 
 can be defined as the size of its core,  the region  where $H(r)$ and  $F(r)$ deviate from one significantly.  This radius is set  by the  Compton wavelength of the gauge boson 
 $R_{mon} = m_v^{-1} = (ev)^{-1}$. 
 The mass of the monopole is 
 \begin{equation} \label{mass}
 M_{mon} \sim {m_v \over e^2} \,
 \end{equation} 
 The proportionality coefficient is equal to $4\pi$
  in the so-called   Bogomolny-Prasad-Sommerfield
 (BPS) limit \cite{BPS1, BPS2}, $h=0$, and is order-one otherwise.  
  The magnetic charge of the monopole 
 is $q_{m} = {1 \over e}$. 
  Naturally, this charge satisfies  the Dirac's charge quantization condition.

  Now, as was already discussed, in order to store the quantum information efficiently, the  object  (in the present case,  the monopole) must support  gapless quantum excitations. Then, such excitations can assume the role of the memory modes.  In the above simplest model the monopole  supports only few  gapless excitations.
   In particular, these include 
   the translation moduli, which represent 
  Goldstone modes of spontaneously broken 
  translation symmetries.  Although their number is 
  not nearly sufficient for making the monopole degeneracy close  to saturation, the scale of Poincare breaking 
  $f$ does play the crucial role  in imposing the entropy bound (\ref{Area}).  This scale, is given by 
  \begin{equation}  \label{fscale}
    f^2 = \frac{M_v^2}{\alpha} = 4\pi v^2  
  \end{equation}   
      
   The paper \cite{Dvali:2019jjw} proposes the two distinct mechanisms 
  for endowing the monopole with a large number of the gapless memory modes that can bring the monopole micro-state entropy 
 close to saturation of the bounds (\ref{Area}), (\ref{Alpha}) 
 and (\ref{Nbound}).  
  These mechanisms incorporate  fermionic or bosonic zero modes.   We discuss the phenomenon
of the swift memory burden first for the fermionic memory modes. 

   \subsection{Memory burden effect with fermion 
   memory modes}

    As the first step,  following  \cite{Dvali:2019jjw}, we shall endow the monopole 
  with a maximal micristate entropy  through the localization of the fermionic zero modes. These modes shall serve 
  as the information-carrier memory modes, 
promoting the monopole into a device of a maximal 
information-storing efficiency.
  
 According to the  index theorem \cite{index},  
a fermion that gets its mass from the Yukawa coupling 
with the Higgs field  $\Phi^a$, results in
a fermionic zero mode localized in the monopole core
\cite{JR}.  
 
  For definiteness, as in  \cite{Dvali:2019jjw},  we introduce
 two multiplets of real Majorana fermions $\psi_{\alpha}^a, 
 \lambda_{\alpha}^a$ which transform as triplets under the gauge $SO(3)$ group. At the same time, they also transform as $N$-dimensional vector representations of some global 
  $SO(N)$-flavor symmetry group.  Here, $a=1,2,3$ and $\alpha = 1,2,.....N$ denote 
  $SO(3)$ and $SO(N)$ indexes respectively.  
  The fermionic part of the Lagrangian has the following form: 
 \begin{eqnarray}   \label{sigma} 
    &&  L_{\sigma} = {1 \over 2} \bar{\psi}_{\alpha}^a
    \gamma^{\mu}D_{\mu}
   \psi_{\alpha}^a \, + \,  
   {1 \over 2} \bar{\lambda}_{\alpha}^a
    \gamma^{\mu}D_{\mu}
   \lambda_{\alpha}^a 
       -  \\ \nonumber
   && - 
   g \epsilon^{abc}\Phi^a  \bar{\psi}_{\alpha}^b \lambda_{\alpha}^c \,,     
 \end{eqnarray} 
 where $g$ is  a dimensionless coupling constant and  
 we use real  $\gamma^{\mu}$-matrixes.  
 
 It is very important to note that the validity 
 of the QFT description is constrained  by the following relations \cite{Dvali:2019jjw}
  \begin{equation} \label{NboundF}
    g^2N \lesssim 1,\, ~  \, e^2N \lesssim 1\, .
  \end{equation} 
  Beyond this bounds, none of  the fields 
  (gauge, Higgs and fermions) represent valid degrees of freedom, and the theory undergoes a  regime-change. 
  This breakdown  is signalled by various symptoms, 
  such as the breakdown of the loop-expansion \cite{Dvali:2019jjw} as well as the saturation of unitarity by scattering amplitudes 
  \cite{Dvali:2020wqi}.  The relation (\ref{NboundF}) plays  a crucial role in constraining the microstate entropy of the monopole by (\ref{Alpha}), (\ref{Area}) and (\ref{Number}).

   Next, due to non-zero Higgs VEV (\ref{Higgsvacuum}) the 
 fermions $\psi_{\alpha}^1, \lambda_{\alpha}^2$ and 
  $\psi_{\alpha}^2, \lambda_{\alpha}^1$ form
  the Dirac fermions with the masses $m_f = gv$ 
 for all values of $\alpha$, whereas the pairs  $\psi_{\alpha}^3, \lambda_{\alpha}^3$ remains massless.  \\  
    
   Now, in the monopole background there exist the fermionic zero modes localized within the monopole core.  
  For $h=0$ and $e=g$ the solution 
  (up to an over-all finite normalization constant)
  has  the form 
  \begin{eqnarray}   \label{sigma} 
    &&  \lambda_{\alpha}^a = {1\over 2} F_{\mu\nu}^a\sigma^{\mu\nu} \epsilon_{\alpha}  
         \\ \nonumber
   &&
   \psi_{\alpha}^a = \gamma^{\mu} D_{\mu} \Phi^a \epsilon_{\alpha} \,,     
 \end{eqnarray}  
   where $\epsilon_{\alpha}$ ($\alpha = 1,2,...N$) are the constant spinors, whereas the bosonic fields are given by the monopole  solution  (\ref{Mon}). 
   The localization radius of  fermionic zero modes is given by, 
   \begin{equation} \label{Fradius} 
     R = (gv)^{-1} = m_{f}^{-1}\,.
   \end{equation}

   These, gapless fermionic modes serve as the 
 memory modes  which store quantum information. 
 The basic memory patterns (\ref{pattern}) are defined 
 by the sequences of occupation numbers $n_{\alpha}$  that can take values $0$ or $1$.   Since fermions are 
gapless,  the patterns are degenerate in energy \footnote{
Of course, the gap is defined up to a precision
$\sim 1/(N R)$  which is the minimal uncertainty 
due to a the spread of the monopole wave-packet.}. 
 Correspondingly, the set of patterns defines the Hilbert space of the monopole microstates of dimensionality 
 $n_{st} = 2^{N}$.   This defines the monopole memory space. 
 The corresponding microstate 
   entropy is, 
   \begin{equation} \label{Smon} 
   S_{mon} = \ln(n_{st}) \sim N \,.
   \end{equation} 
  As already pointed out in \cite{Dvali:2019jjw},  
   for the maximal degeneracy permitted by 
  (\ref{NboundF}), the monopole entropy  saturates all three bounds  (\ref{Area}), (\ref{Alpha}) and (\ref{Number}). 
  Correspondingly,  such a monopole has a maximal efficiency of information storage. In other words,  it  is a ``saturon".    
  
  In order to quantify  this efficiency,    
  first notice that the energy difference between an
   information pattern stored in the $N_p$-excited 
    monopole zero modes  and the identical pattern stored 
   in a wave-packet of free fermionic modes is, 
    \begin{equation} \label{Evac} 
   E_p|_{r \gg  R} = N_p gv = N_p \frac{1}{R} \,.
   \end{equation}  
   This energy-difference originates from the fact that 
   the fermions in the asymptotic vacuum are gapped, with masses 
 $gv = 1/R$.   For a typical memory pattern this energy difference is macroscopic.   
 In particular, for $N_p \sim N$ it becomes comparable to the  mass of a monopole.   
 
  Secondly, 
 the energy difference between the two patterns 
 $N_p$ and $N_p'$  in the asymptotic vacuum is 
   \begin{equation} \label{DeltaE} 
   \Delta E_{pp'}|_{r \gg R} = (N_p-N_p') gv = (N_p-N_p')\frac{1}{R} \,.
   \end{equation}  
 The above relations demonstrate  that due to fermionic zero modes the saturated monopole represents a device of maximal efficiency of information storage. 
  In this sense, the monopole reproduces 
  the features of the black hole entropy (\ref{BekH})  
  with the substitution
  $v = M_P$.

 \subsection{A swift activation of the fermionic memory burden} 
 
   Let us consider a monopole with a memory pattern 
  with $N_p$ excited fermionic zero modes. 
   In the monopole background, this information pattern 
   costs almost no additional energy as compared to the monopole 
   ground state. Equivalently, we can say that the monopole ground state is highly degenerate.   Of course, 
   the pattern cannot be extracted due to the energy barrier created by the asymptotic energy cost 
   (\ref{Evac}) of the same pattern. 
   
    However, the stability of the monopole is due to its topological charge rather than the information content. 
    In other words, the monopole experiences almost no 
    memory burden from the information it carries.
     The situation changes dramatically if the monopole 
  is subjected to  an external disturbance that affects the 
  gaps of the memory modes.  In response to such an external stimulus, the memory burden gets 
activated and strongly affects the dynamics. 

 As a particular example of such disturbance, let us consider the merger of a monopole with an anti-monopole. 
   We assume that the initial separation of monopoles 
 to be large $r_{in}  \gg R$.  
Since their magnetic charges are opposite,  the two object experience a magnetic attraction. In addition they experience the attraction due to the Higgs force, 
which for $h=0$  has the same strength as the magnetic one.   In BPS limit, for monopole-monopole the two forces 
cancel out \cite{BPS3}, whereas for monopole-anti-monopole they add.  
 Correspondingly, there is a Newtonian-type attractive potential between the two monopoles given by, 
  \begin{equation} 
    V(r) \, \sim  \, - \frac{1}{e^2}\frac{1}{r}  \,. 
  \end{equation}   
      Under the influence of this attraction, the monopoles 
      fall towards each other.

      The dynamics of the monopole-anti-monopole   scattering  and annihilation, without taking 
      into account the memory burden effect, 
       has been studied numerically in number of articles
    \cite{MMbar1, MMbar2}.  Analogous studies 
    for confined monopole-anti-monopole pairs have also 
   been performed (see, \cite{MMbarC} and references therein).      
    
    The point we wish bring across is that 
    for monopoles endowed with fermionic memory patterns,   
    this classical dynamics gets affected by the swift memory burden effect. 
    
     For definiteness, let us endow only one of the  monopoles 
     by the memory pattern with a memory-load $N_p$.   Initially, when the anti-monopole is far away, the memory modes localized on the monopole   do not feel 
 its presence. This is because the profile functions 
 of the zero mode fermions are not affected. 
    However, as the distance between the monopole and the 
    anti-monopole shortens, 
    the gaps of the memory modes get affected. 
    In particular, at separation  $r \sim R$, the 
    gaps become of order $1/R$.  The cost of the memory patterns at this point becomes, 
      \begin{equation} \label{Enear} 
   E_p|_{r \sim  R} \,  \sim \, N_p\,  gv \, = \, N_p\frac{1}{R} \,.
   \end{equation}  
   The corresponding memory-burden parameter  
   is, 
       \begin{equation} \label{muMon} 
   \mu \,  \sim \, \frac{M_{mon}} { 
   N_p\,gv}   \,.
   \end{equation}     
 For a pattern with maximal memory load,  $N_p \sim N$, the  vacuum energy cost of the memory pattern 
 (\ref{Enear}) is of the order of the monopole mass. 
    Therefore, the  memory burden parameter is order-one, 
         \begin{equation} \label{MuMonmin} 
   \mu \,  \sim \, \frac{M_{mon}} { 
   N\,gv}  \sim 1 \,.
   \end{equation}         
     Correspondingly, the dynamics 
    of the monopole merger will be affected 
    substantially. 
    
  Monopoles cannot annihilate prior to getting rid 
 of the  information pattern stored in fermions. 
 However, since  the global $SO(N)$-charge must be conserved, the information must be radiated  away 
 in form of the the bulk fermions.  The least, this delays the annihilation process and affects the radiation spectrum 
 in wavelengths $\sim R$.  

  It is important to avoid a false impression that the 
  strength of the memory burden is proportional to 
  a conserved $SO(N)$-charge. This is not the case.  
  The swift memory burden will take place even  if the total $SO(N)$-charge carried by the system is zero.  As long as the memory pattern carried by  the system is not empty, 
  the dynamics shall be affected.

  \subsection{Nambu-Goldstone memory modes} 
  
 Let us now consider the sudden memory burden effect 
 due to localized bosonic  memory modes in the monopole background. 
  An  explicit model that gives rise to such zero modes, 
   which can endow the monopole with maximal memory-storage capacity, was already introduced in 
   \cite{Dvali:2019jjw}.  We shall first briefly review the model 
   and then discuss the memory burden effect. 
   
   The memory modes emerge as the gapless Goldstone excitations of  a global $SO(N)$ flavor symmetry, which is spontaneously broken in the monopole core.  
   At the same time, the symmetry is unbroken in the asymptotic 
   vacuum. Correspondingly, the Goldstone modes are 
   strictly localized within the monopole core.  
   
  In order to achieve this, we couple the monopole field 
  to a real scalar  
   $\sigma_{\alpha}, \alpha = 1,2,...N$  transforming as $N$-dimensional vector representation
 of $SO(N)$.      
   Of course, both the group  as well as the representation content are chosen for definiteness. The mechanism 
   is operative for other groups (e.g., $SU(N)$)  as well as other representations.  
    
    Repeating the construction of  \cite{Dvali:2019jjw}, we  add the following terms to the Lagrangian 
   \begin{eqnarray}   \label{sigma} 
    &&  L_{\sigma} = {1 \over 2} \partial_{\mu}\sigma_{\alpha} 
  \partial^{\mu}\sigma_{\alpha} \, - \, 
  {1\over 4} g_{\sigma}^2 ( \sigma_{\alpha}\sigma_{\alpha})^2 
     -  \\ \nonumber
   && - {1 \over 2}(
   g^2 \Phi^a\Phi^a  - m^2)  
   ( \sigma_{\alpha}\sigma_{\alpha}) \,,     
 \end{eqnarray}  
  where, $m^2> 0$ is a mass parameter and
  $g^2>0$ and $g_{\sigma}^2 >0$ are dimensionless 
  coupling constants. 
  The validity of the QFT description  puts the following bound on the  parameters of the theory,   
  \begin{equation} \label{Nbound}
    g^2N \lesssim 1,\, ~  \, g_{\sigma}^2N \lesssim 1\, .
  \end{equation}
 
 Now, on the monopole background  
 (\ref{Mon}), the $\sigma$-field acquires  an effective $r$-dependent 
mass term
\begin{equation}\label{massr} 
m^2(r) =  g^2 v^2 H(r) - m^2 \,. 
\end{equation} 
   Since we want the global symmetry to be unbroken in the vacuum,   we choose the asymptotic value of the mass-term to be positive   
 $m^2(\infty) =  g^2 v^2 - m^2 > 0$.  
With this choice, the VEV of the $\sigma$-field vanishes 
away from the monopole $r \rightarrow \infty$
 and there exist no gapless excitations among the asymptotic modes. 
  
 However,  the effective mass (\ref{massr})  becomes imaginary  in the monopole core.  This signals 
 a potential instability which may force $\sigma$ to develop 
 a non-zero expectation value in the monopole core. 
 However, the outcome depends on a detailed balance 
 between potential and gradient energies.  
   
  It was shown in \cite{Dvali:2019jjw} that there 
exist a parameter range for which  
it is energetically favorable for $\sigma$ to condense inside the monopole. 
In particular, the conditions are 
\begin{equation}\label{eg1}
 g/e \gtrsim 1\,. 
 \end{equation}
 Under these conditions,
  the ground-state of the system is described by a function 
  $\sigma_{\alpha} (r)$ that has a non-zero expectation value in the monopole core,  $\sigma_{\alpha} (0) \neq 0$. 
    Without any loss of generality, we can choose 
    the basis as $\sigma_{\alpha} (r) = \delta_{\alpha 1} \sigma(r)$.
  
     A detailed energetic analysis shows (see, \cite{Dvali:2019jjw} ) 
that  the localization  radius $R$ of the $\sigma$-condensate and its value in the center of the monopole
are given by 
    \begin{equation} \label{scales1}
    R \sim (ev)^{-1},\, ~  \, \sigma(0) \sim {g \over g_{\sigma}} v\,.
  \end{equation} 
    At the same time, the requirement that the back-reaction from the condensate 
  $\sigma(r)$ to the monopole solution (\ref{Mon})  is weak, implies 
     \begin{equation} \label{scales2}
    g^2  \lesssim g_{\sigma} e\,.
  \end{equation} 
In this case the correction to the monopole mass due to a 
non-trivial profile of the $\sigma$-field is small and the total energy of the configuration is well-approximated  by (\ref{mass}).  

 Now, the non-zero VEV of $\sigma_{\alpha}$  in 
 the monopole core  breaks the $SO(N)$-symmetry  spontaneously  down to $SO(N-1)$.    
 Correspondingly, there exists $N-1$ 
 gapless Goldstone bosons.  
  They correspond to local transformations 
  of the  $\sigma_{\alpha}$-VEV by the broken 
  $SO(N)$-rotations, $O_{\alpha \beta}(x) \sigma_{\beta}(r)$, where $O \equiv {\rm e}^{i\theta^AT^A}$. 
  The effective (one-dimensional) world-volume action 
  of the Goldstone modes has the following form, 
    \begin{equation}\label{Agold}  
  S_{\theta}  =\int dt \, {\mathcal N_{\theta}} (\dot{O}_{\alpha, 1})^2 \,,
  \end{equation} 
where, 
    \begin{equation}\label{Ngold}  
  {\mathcal N_{\theta}} \equiv  4\pi \int_0^{\infty}  r^2 dr \sigma^2(r)
   \,, 
  \end{equation} 
  is the Goldstone norm. 
 In a linearized approximation we  have, 
    \begin{equation}\label{Agold}  
  S_{\theta}  =\int dt \, {\mathcal N_{\theta}} \, \sum_{A} \,  (\dot{\theta}^A)^2 \,,
  \end{equation} 
  where the sum in $A$ runs over the broken generators.
 
  The excitations of the gapless Goldstone modes 
  create a large number of the degenerate microstates of the monopole.   Their number can be found by counting the 
  number of the degenerate microstates in  a quantum theory of $N$ gapless oscillators,   
 with creation/annihilation 
 operators that satisfy the usual algebra, 
 $[\hat{a}_{\alpha}, \hat{a}_{\beta}^{\dagger}] = \delta_{\alpha\beta}$, 
 subject to the following constraint on their occupation numbers 
 \begin{equation}\label{GoldVac}  
  \sum_{\alpha = 1}^N \hat{n}_{\alpha}  = N_{\sigma}  \, .    
  \end{equation} 
  Here,  $\hat{n}_{\alpha} \equiv \hat{a}_{\alpha}^{\dagger} \hat{a}_{\alpha} $ are the number operators and 
   \begin{equation}\label{Nsigma}  
  N_{\sigma} \sim  {g^3 \over g_{\sigma}^2e^3}  \, , \end{equation} 
   is the average occupation number of quanta in
   the $\sigma$-condensate.  The number of the degenerate microstates is  given by the following binomial coefficient 
  \begin{equation}\label{Nstates} 
 n_{st} =  \begin{pmatrix}
    N_{\sigma} + N -1   \\
     N_{\sigma}   
\end{pmatrix} \,,
  \end{equation} 
  
 Now, taking the saturation point of the bound (\ref{Nbound}) 
 and using the relations  (\ref{eg1}), (\ref{scales2}) and (\ref{Nsigma}), we arrive to the following 
  limiting expressions, 
   \begin{equation} \label{Nsaturation}
    {1\over g^2} \sim {1 \over e^2} \sim {1 \over g_{\sigma}^2} \sim 
    N\, \sim N_{\sigma} .
  \end{equation} 
 Thus, all dimensionless couplings 
 take the common value that can be denoted by an 
 universal  coupling $\equiv \alpha \sim g^2, e^2, g_{\sigma}^2$, which satisfies, 
     \begin{equation} \label{alphaNN}
   \alpha \sim  \frac{1}{N}.
  \end{equation} 
  
 Next, evaluating the equation (\ref{Nstates}) for $N = N_{\sigma}$ and using Stirling's approximation, we get the following 
number of microstates,  
 \begin{equation} \label{nstM}
 n_{st} \sim  \, {\rm e}^N \,.
 \end{equation} 
 The corresponding micro-state entropy of the monopole scales as, 
 \begin{equation}\label{SsatMon} 
 S_{mon} \equiv  \ln (n_{st}) \sim N  \, .
  \end{equation}

  Now, using the relations  (\ref{scales1}) and  (\ref{mass}),
 the above maximal entropy of the monopole can be written 
 as,  
  \begin{equation} \label{SSSmon}
    S_{mon}  = N = (R_{mon}v)^2 = \frac{1}{\alpha} \,.   
  \end{equation} 
  We thus see that the monopole entropy at the 
saturation point reproduces the bounds (\ref{Area}),
(\ref{Alpha}) and (\ref{Nbound}).    
  
  This reproduces the result of  \cite{Dvali:2019jjw}, 
  showing  that 
 the Goldstone zero modes endow the monopole with 
 maximal efficiency of information storage and
 the entropy of saturated monopole
 (\ref{SSSmon})  is identical to  the Bekenstein-Hawking 
 entropy of  a black hole (\ref{BekH}) with the substitution 
  $v \, \rightarrow \, M_P$. 

  As already explained, this connection between the scales 
  has a very clear physical meaning, since  
  the parameters  $v$ and $M_P$ represent the 
  scales of spontaneous breaking of Poincare symmetry by the monopole and the black hole, respectively.     
  
  We must note that for finite $N$, the localized Goldstones are not exactly gapless but have the frequency gaps
 \begin{equation} \label{minGap}
\epsilon_{\rm min} \sim \frac{1}{N\, R} \sim \frac{1}{S\,R}\, .
 \end{equation} 
  The reason is that the monopole represents 
 a localized wave-packet which spreads over time 
 $t \sim RN$.  This spread introduces a fundamental 
 spread of the energy levels. In the language of spontaneous symmetry breaking, the small non-zero energy gaps of Goldstones can be understood from the fact that 
 at finite $N$, the Hilbert spaces corresponding to  different 
 orientations of the order parameter are not exactly orthogonal.  Presence of the elementary 
gap is universal and is equally shared by fermionic 
zero modes. 

Of course, as already discussed, the same feature is also present 
 in black holes.    Indeed, the equation (\ref{minGap}) is identical  to (\ref{Lspread}) which describes the 
 effective gaps of the black hole memory modes 
 induced by the spread of the wave-packet.

 The microscopic spread in the memory mode frequencies  does not affect the entropy count, since the entire set of  microstates fits within the energy gap $\sim 1/R$.  For a black hole, this is the energy of a single Hawking quantum.   For a black hole and other unstable saturons
 \cite{Dvali:2021rlf, Contri:2025eod, Dvali:2021tez},
  the spread  (\ref{minGap}) also matches the level-width created due to the decay. 
 However, it is important to understand that even for stable 
 saturons, such as the monopole, the minimal gap 
  (\ref{minGap}) is unavoidable due to the spread of the wave-packet.

   We  are now ready to study the swift memory burden effect  due to the Goldstone modes. 
     As in fermionic case, we consider a situation in which the monopole  merges with an anti-monopole.
      We study the two manifestations of the memory burden effects. 
      
   \subsection{The memory burden from misalignment 
   of memory patterns. } 
   
      We consider monopole and anti-monopole 
      with misaligned memory patterns. 
      That is, we assume that the VEVs of the
      field $\sigma_a$ in the two locations are  relatively rotated  by an angle $\Delta \theta$ in one of the Goldstone 
      directions.  When the monopole and anti-monopole are
      far apart, such a rotation does not cost any energy since 
       $\sigma(r)=0$ in the intermediate vacuum.  
       Correspondingly, the memory space of separated monopole and anti-monopole represents a direct 
       product of the two memory spaces, without any cross-coupling. 
       
       However, when the two objects approach each other at a distance smaller than the localization radius 
       of the sigma VEV, the two patterns start  to overlap
       and create an additional energy cost. The gradient energy due to Goldstone misalignment  has the form
      \begin{equation} \label{EdeltaPM1}
      E_{MB} =  \int d^3\vec{x} \sigma(x)^2 \left (\dot{\theta}^2
      +  (\vec{\nabla} \theta)^2 \right ) \,, 
      \end{equation}    
     Which can be estimated as, 
    \begin{equation} \label{EdeltaPM}
      E_{MB} \sim {\mathcal N}_{\theta}\left (\frac{\Delta \theta}{R} \right )^2\,.
    \end{equation}     
 
 Notice that for a monopole that saturates the entropy 
 bound,  for $\Delta \theta \sim 1$, the above energy is comparable to the mass of the monopole, 
      \begin{equation} \label{EdeltaPM}
      E_{MB} \sim  M_{mon}\,.
    \end{equation}     
 This implies that, upon a merger, an order-one fraction 
 of the initial energy gets converted into 
 the gradient energy of the misaligned memory patters.
 This creates a swift memory burden effect which  influences the merger process as well as the subsequent radiation.  The detailed quantification of the effect 
 requires a separate analysis which will not be performed here.  However, the qualitative effect is clear. 
   The important thing is that the effect of the memory burden is macroscopic and is imprinted into the radiation pattern.     
     
  \subsection{Memory burden from excited Goldstone memory modes} 
  
  Let us now consider a situation when the memory pattern 
  is stored in set of excited Goldstone modes
of non-zero frequencies.  
   For example, we can spin the VEV of $\sigma_{\alpha}$ 
in internal space by one of the broken generators, 
 \begin{equation}   \label{spint}
    \sigma_{\alpha}(r,t) =  \sigma(r) \left(\delta_{\alpha 1} \cos(\omega t) + \delta_{\alpha 2} \sin(\omega t) \right) \,.
    \end{equation}  
     This anzats effectively endows the monopole 
     with the $SO(N)$-charge: 
   \begin{equation}   \label{charge}
   Q  = 4\pi \int r^2 dr \sigma(r)^2 \omega = 
   {\mathcal N_{\theta}}  \omega \,.
 \end{equation}             
 The modulus function $\sigma(r)$ satisfies the following equation,  
  \begin{equation} \label{radial}
\mathrm{d}_r^2\sigma(r) + \dfrac{2}{r} \mathrm{d}_r\sigma(r) +   (\omega^2 -  m^2(r))\sigma(r)  - g_{\sigma}^2 \sigma^3(r)  
=0 \,,
\end{equation}
 where $m(r)$ is given by (\ref{massr}). 
 
   We already know that for $\omega = 0$
   a non-trivial solution   with $\sigma(0) \neq 0$ exists for 
   a finite range of the parameter $m(\infty) = m$.
    From the point of view of the above equation, 
   the effect of  $\omega$ is to shift this value. It is therefore obvious that we can maintain the same solution 
   for a shifted value  $m^2 \rightarrow m^2 - \omega^2$.   
   Correspondingly, the solution with non-zero charge 
exists for a finite range of the parameter $\omega$. 

 In the sense of carrying the $SO(N)$-charge,  
this solution shares some features with the 
memory-burdened vacuum 
bubble solution  obtained in \cite{Dvali:2021tez, Dvali:2024hsb}.  There the memory burden was due to a Goldstone charge. 
 
     However, an important difference from the construction of \cite{Dvali:2021tez, Dvali:2024hsb} is that in 
     the present case  the solution with non-zero charge exists only on top of the monopole background. 
  That is, the monopole is not stabilized by the memory burden effect  but solely by its topology.    That is, for an isolated monopole  the memory burden is ``dormant". 
  This aspect is similar to the situation with an unperturbed 
classical black hole, which is also stable regardless of the 
information  load it carries.

However, the burden  shall get activated swiftly if the monopole meets an anti-monopole.   The effect exists for 
arbitrary values of charges but with different outcomes. 
For example, we can assume that the  Goldstone charge 
  of the anti-monopole is zero. 
  
   The fact that the Goldstone charge will lead to a swift memory burden response is obvious from the fact that if monopoles annihilate
   the charge has to be released in the vacuum. 
   This is costly in energy, since  in the vacuum without the monopole support, the quanta of the $\sigma$-field are highly gapped.  Correspondingly, the merger process will be altered 
   macroscopically. In particular, the process of annihilation  of monopole with an anti-monopole is expected 
  to be prolonged.   Whether the swift memory burden created by the Goldstone charge can prevent a full annihilation and create a stable bound-state of monopole 
  and anti-monopole   is a dynamical question that requires more detailed  analysis.  However, as in the case of a black hole,  for a maximally loaded memory pattern, the energy balance of the process is affected by  the amount comparable to the monopole mass. 
  
  Analogous analysis of the swift memory burden effect 
 can be performed for monopole-monopole scattering,  
  which previously has been studied in the absence of the memory burden \cite{MM1, MM2, MM3} (for recent numerical analysis see, \cite{MM4}).

 \subsection{Similarities and differences 
 in swift memory burdens in black holes 
 and solitons }   
 
  There are clear similarities between black holes and 
  solitons subjected to the swift memory burden 
  effect.  This is logical in the light of  
  presence of the memory burden effect in saturated 
  solitons \cite{Dvali:2021tez, Dvali:2024hsb}.
     
    In the present case, the monopoles are the analogs of the merging 
    black holes, with the identification $v = M_P$.  
    The fermionic or bosonic zero modes  correspond to the black hole memory modes. 
    
   However, there are clear differences, in particular, 
   due to the nature of the memory modes in the two systems. These differences suggest that the memory burden effect in black holes should be more dramatic.

    In the case of a monopole, the largest 
   energy gap between a memory mode and the corresponding mode in the vacuum is $\sim 1/R$. 
   In contrast,  for a black hole the  analogous gap can be much larger, $\sim M_P \sim  \sqrt{S}/R$. 
     Due to this difference, from the point of view of the memory burden effect, 
  in the classification of \cite{Dvali:2024hsb}, a  saturated  monopole  and 
  a black hole belong to the type-$I$ and type-$II$ systems respectively.

    Correspondingly,  the difference between the energy costs 
    of a typical maximal information pattern in and outside of a black hole  exceeds the mass of a black hole by a factor 
   $\sqrt{S}$.  In contrast, the analogous  energy difference in  case of a saturated monopole is  order-one.  This
  difference  is  also quantified by the 
   difference between the memory burden parameters 
   for the maximal information loads in the two systems,
   which are  given by (\ref{minMu}) and   (\ref{MuMonmin}) respectively.    
   
     Another difference is that, in  case of a monopole-anti-monopole example,  the end result of a classical collision can be a total annihilation.  The system then is left with no gapless memory modes. In contrast, in case of a black hole merger the number of the available memory modes always increases as compared to the initial state. 
    This however is not an essential difference, since the solitonic systems  with a similar behaviour can readily be constructed.   For example, in the present case one  can consider a monopole-monopole merger, which of course does not lead to any annihilation.  
    We notice that various mergers of the saturated vacuum bubbles  have been studied numerically \cite{Dvali:2023qlk,  MichaelZ}. These can be adopted for the simulations of the swift memory burden effect.  
      
    \section{Laboratory tests of memory burden effect} 
 
  In this chapter, we would like to outline some 
 proposals for the study of the memory burden effect 
 in systems  that are potentially accessible in table top laboratories.  
  The idea is to create a simple system exhibiting the phenomenon of the assisted gaplessness and bring it to the critical state  in which  the memory modes become nearly gapless.  Next,   one encodes an information pattern in the excitations of these modes. The memory burden effect will manifest itself as a resistance against the removal of the system from this critical point. 
   
   In particular, the quantum depletion 
   of the system will be suppressed by the memory-load. 
    This  shall serve as an analog of the quantum memory burden stabilizing the system against the Hawking evaporation.    
    
    The system will also exhibit  a swift memory burden effect   that will affect its classical response to various perturbations.   
       
     By now, the identification of the prototype systems 
     has already been achieved.  In fact, the original 
    proposal     
    of the memory burden effect  \cite{Dvali:2018xpy}   
    was performed in the QFT Hamiltonians that are ready-made for the many-body implementations in the quantum labs.
 
 Such are the systems with attractive cold bosons.  In particular, 
 it has been shown  \cite{Dvali:2017nis}
  that  if the attractive interaction is (angular) momentum dependent,   a single quantum field suffices for producing a degenerate set of the gapless memory modes that 
 provide the area-law microstate entropy, strikingly similar 
 to a black hole.    
   
 However,  as it was already discussed, the memory   burden effect is exhibited already by  a minimal system with the assisted gaplessness, without the need of a large number of the memory modes. Therefore, for the purpose of describing the essence of the  experimental tests, it is sufficient to discuss such systems.  
   
  We shall  consider a simple many-body model 
  with $N$ attractive bosons (e.g., atoms) on a ring
  \cite{ring, Kanamoto:2008zz}.
  This model was used in  \cite{Dvali:2012en}  as a simple prototype model describing the essence of criticality of the graviton condensate within the microscopic theory of black hole's  $N$-portrait.  
   
  Various aspects of this model were further discussed 
  in the series of papers, \cite{Flassig:2012re, Dvali:2013vxa,  Dvali:2015ywa, Dvali:2015wca, Dvali:2016zqx}.
   These studies revealed that, at least at the qualitative level, the model captures  many aspects  of black hole information processing predicted by the $N$-portrait.        
   The version of the model with non-periodic boundary conditions has also been studied in \cite{Dvali:2018tqi}. 
    However, in the discussion below, we shall stick to 
 the original periodic case.        
  
  The  Hamiltonian on one-dimensional ring of radius $R$ is 
       \begin{eqnarray}  \label{LH}
  { \mathcal H}\,  =  \, &&   \int dx  \, 
 \hat{\Phi}^{\dagger} \frac{\hbar^2 \Delta }{2M}  \hat{\Phi }  
  - g\hbar \, \hat{\Phi}^{\dagger}  \hat{\Phi}^{\dagger}  \hat{\Phi} \hat{\Phi} \,, 
 \end{eqnarray}
  where $\hat{\Phi} = \sum_{\vec{k}} \frac{1}{\sqrt{2\pi R}} e^{i \frac{\vec{k} \vec{x}}{R}} \hat{a}_{\vec{k}}$ is a non-relativistic bosonic field
  operator,  with creation/annihilation operators 
  of momentum modes satisfying 
  the  usual algebra $[\hat{a}_{\vec{k}}, \hat{a}_{\vec{k'}}^{\dagger} ]= 
   \delta_{kk'}$.  $M$ is the mass of the boson and  
   $g > 0$ is a coupling constant. 
   
    The Hamiltonian written in terms of the mode operators is
  \begin{equation}
  {\mathcal H}\,  =  \,   \epsilon_0   \left (
  \sum_{k =-\infty}^{+\infty} k^2 \hat{a}_k^{\dagger}\hat{a}_k 
  -  \frac{\alpha}{4} \sum_{k_1+k_2-k_3-k_4 =0} \hat{a}_{k_1}^{\dagger}\hat{a}_{k_2}^{\dagger} \hat{a}_{k_3}\hat{a}_{k_4}  \right ) \,,
 \end{equation} 
 where $\alpha \equiv \frac{gm}{\pi \hbar}$ is a dimensionless coupling and $\epsilon_0 \equiv 
 \frac{\hbar^2}{2R^2M}$ is an elementary energy gap setting the 
 spacing between the various momentum modes 
 for a non-interacting part of the Hamiltonian.

   We study the system around the state in which the field 
   is macroscopically occupied to a certain large number  $N$.  We choose the macroscopically occupied mode to 
 be $k=0$ and study the spectrum of excitations 
   around the state $n_0 = N$.  We limit ourselves  
   with the excitations of modes with $k =\pm 1$.  
 
Notice that the above system represents an example of the 
 Hamiltonian (\ref{HintG}), in which the role of the master mode is played by the zero-momentum mode 
 $\hat{a}_0$.  
     
      Performing the Bogoliubov transformations  
    \begin{equation}  \label{Bmodes} 
   \hat{a}_{\pm1} = u_{\pm}  \hat{b}_{\pm1} + v_{\pm}^* \hat{b}^{\dagger}_{\mp1}
   \end{equation}               
with 
\begin{eqnarray} \label{Efunction}
&&u_{\pm}^2 = \frac{1}{2} \left ( 1 + 
\frac{1- \alpha N/2}{\epsilon} \right ),~
 v_{\pm}^2= \frac{1}{2} \left ( - 1 + 
\frac{1- \alpha N/2}{\epsilon} \right ),  \nonumber
\\ 
&& {\rm where} \, ~~~~~~\epsilon \equiv (1-\alpha N)^{\frac{1}{2}}\,,
\end{eqnarray}         
     we get the following 
       effective Hamiltonian for the Bogoliubov modes \cite{Dvali:2016zqx} : 
                \begin{eqnarray}  \label{LHb}
  {\mathcal H}\,  =  \, &&   \epsilon_0 \left ( 
  \epsilon 
  \sum_{\pm 1}  \hat{b}_{\pm 1}^{\dagger}\hat{b}_{\pm 1}
  -  \frac{1}{N\epsilon^2} {\mathcal O} (\hat{b}^4)\, \right ) 
 \end{eqnarray}
       
        The above system represents an example of 
      the assisted gaplessness, with 
      $\hat{b}_{\pm 1}$ modes playing the role of the 
      memory modes and the zero mode $\hat{a}_0$ 
      playing the role of the master mode. 
    Indeed,  taking into account  (\ref{Efunction}), it is clear that  (\ref{LHb}) realizes a version of the generic 
  Hamiltonian (\ref{HintGB}) with the   
  critical exponent of the gap function 
  $p=\frac{1}{2}$.     
      
      The gaps of the $\hat{b}_{\pm 1}$-modes can be made arbitrarily small  by taking a double scaling limit  $\epsilon 
      \rightarrow 0$ and   $N \rightarrow \infty$ 
    while keeping  $\epsilon^3 N=$fixed.    
   This is because, within the validity of the 
   Bogoliubov Hamiltonian,  the occupation number of the memory modes is  bounded from above by $n_{\rm max} \sim \epsilon^3 N$. 
        
      Now, we can store information in the excitations of the 
     nearly-gapless Bogoliubov modes. 
      The information pattern $\ket{n_+,n_-}$
       is represented by the  occupation numbers $n_+$ and $n_-$. 
  This creates the memory burden effect in form of the 
     energy barrier that stabilizes  the master mode 
     $\hat{a}_0$ at the critical point $\alpha N = 1$. 
     
       The decrease of the occupation number 
      of the master mode by the amount
      $\Delta n_0$, causes the energy difference 
      \begin{equation}  \label{BBburden}
      \Delta E \, = \,  \left (\frac{\Delta n_0}{N} \right )^{\frac{1}{2}}
       \epsilon_0 (n_+ + n_-)\,,
     \end{equation} 
     which creates the memory burden effect. 
     This effect will influence the  quantum  as well as the classical  evolutions of the system. 
     
       In particular, the memory burden effect resists to a quantum depletion  of the master mode. This means that the quantum depletion should be observed to take longer. 
       
       Moreover, the swift memory burden effect will take place  as a response  to an attempt of moving the system 
      away from the critical point.  For example,
      the burden creates an additional potential 
      which resists to a change of the size of the  ring, $R$,
    which affects the energy gap $\epsilon_0$. 
      
      In particular, we can make the system overcritical, 
    $\alpha N > 1$,       
    by  decreasing  the radius of the ring.
  In this regime, classically, the ground state of the system corresponds to a localized bright soliton. 
    Correspondingly, the system evolves towards it. 
    The time-evolution without the memory load has 
    been studied in \cite{Dvali:2013vxa}. 
    Naturally, such evolution exhibits no memory burden effect.  
    Namely, in the over-critical regime the system develops a Lyapunov exponent and relaxes towards the soliton ground state. 
    
   In the absence of the memory load, the 
   initial state is an uniform condensate of the zero-momentum modes,  which is translationally-invariant.  
   Correspondingly, it evolves towards  the 
   translationally-invariant ground state which can be viewed as the superposition of solitons  uniformly distributed over the ring.  This state is of course highly entangled. 
   
   Now, the idea of the experiment would be to compare the 
   behaviours of the system with and without the memory loads. 
     First, the memory-load shall make the system 
   more stable. If the load is sufficient, it can even 
   cancel the Lyapunov exponent.  
   
   Secondly, even if the instability persists, 
   the evolution must be  strongly affected by the nature of the memory pattern. For example,  the memory pattern 
   can be prepared in a translationally non-invariant state, if 
   we have $n_+ \neq n_-$.  In this case,  
   translationally  non-invariant  classical 
   trajectories will be selected and the system shall not evolve towards the translationally-invariant ground state. 
    
   It has been shown  \cite{Flassig:2012re,  Dvali:2015wca, Dvali:2016zqx}  that near the critical state the systems develops entanglement. The corresponding  time-scale is macroscopic  and scales as  $t_{\rm ent} \sim  \sqrt{N}$  \cite{Dvali:2015wca}. The time evolution 
   of a perturbed entangled state must also be affected by the memory burden effect. 
   
    Finally, the number of the gapless memory modes can be increased if the field $\hat{\Phi}$ transforms under a larger internal symmetry, or if the interactions are momentum-dependent as in  \cite{Dvali:2017nis, Dvali:2018xpy}.


  \section{Conclusions and outlook} 
  
   In this paper we have introduced a particular 
   manifestation of the memory burden phenomenon 
 \cite{Dvali:2018xpy, Dvali:2018ytn, Dvali:2020wft},  named 
  the ``swift memory burden effect", and argued that it 
  is generic for systems of  high information capacity, such as black holes. 
   Its essence is that the information load carried 
   by a black hole, upon a perturbation, is expected to  significantly influence the subsequent classical dynamics.
   
   That is, a classical black hole possesses a new hidden macroscopic  characteristics in form of the information load, 
  quantified by the memory burden parameter $\mu$ (\ref{MuBH}).  
 In the black hole ground state, the information load carried  by it, is ``dormant".    Due to this, the unperturbed black holes carrying vastly different information loads are degenerate in mass and in other classical characteristics.  
    However, upon a perturbation,  the memory  burden 
    gets activated swiftly.  Correspondingly, the black holes  with different information loads $\mu$ exhibit very different classical dynamics.  Naturally, this can have some important implications, including the observable effects in the black hole spectroscopy. 
    
 The generic memory burden effect introduced in 
 \cite{Dvali:2018xpy, Dvali:2018ytn, Dvali:2020wft}  
  is an universal phenomenon  exhibited 
  by the systems of high energetic efficiency of information storage.  Its essence is that an information load 
  carried by the system tends to stabilize it.   
  Namely, in all such systems, the cost of the information pattern is minimal in the state of the assisted gaplessness where a large diversity of the memory modes are gapless. 
  Correspondingly, in such a state the system exhibits 
  the maximal efficiency of the information storage. 
 The back-reaction from the information load resists 
 to any departure of the system from this state.
 In particular, it stabilizes the system against the decay.  
 The effect is universal and has been demonstrated to exist 
 in generic systems of enhanced information capacity,  including the saturated solitons \cite{Dvali:2021tez, Dvali:2024hsb}.

  Since black holes are the most prominent representatives 
  of such systems, they are likely subjected to 
  the memory burden effect.   In fact, many consistency arguments as well as the microscopic picture indicate that in black holes the phenomenon must be especially 
sound. 
  
  So far, the implications of the memory burden effect have been considered in the context of black hole stabilization against  the Hawking decay
 \cite{Dvali:2018xpy,  Dvali:2020wft,  Dvali:2021tez, Alexandre:2024nuo, Thoss:2024hsr}. 
  In other words, in these studies the focus was on the memory burden phase achieved over a relatively long period of a gradual decay.  Not surprisingly, this effect has number of implications, as it 
 modifies Hawking's semi-classical evaporation regime.

  In the present paper we have pointed out 
 another manifestation of the memory burden effect which does not require a large waiting time.
 Instead, we argued that, regardless of  the elapsed decay time, 
 the memory burden will be activated swiftly 
 whenever the black hole is subjected to
 a significant perturbation.  
 In particular, the role of such perturbation can be assumed  by a merger with another black hole or a star.   
  That is,  a swift memory burden response is caused by essentially any  classical (i.e., quantum-coherent) evolution of a black hole, since any such evolution  is expected to take the memory modes away from the gaplessness.  In particular,  the semi-classical quazi-normal modes  are expected to be influenced by the information load carried by the black hole.

   We have formulated a calculable framework 
 which allows to capture the most essential features
 of the phenomenon and make some general predictions.  Within this framework 
 we have parameterized  the swift memory burden 
 response and derived some master formulas.     
   The strength of the swift memory burden response depends on  the memory burden parameter $\mu$, defined 
   in (\ref{MuM}).  For a black hole of mass $M$ this parameter  is given by (\ref{MuBH}).

    This quantity captures the energetic efficiency  of the given information load carried by a black hole.   The lower bound 
  on $\mu$, expressed in terms of the black hole entropy, is
  \begin{equation} 
     \frac{1}{\sqrt{S}} \lesssim \mu \,. 
     \end{equation} 
   This bound is reached when the information carried by a collapsing source uses the entire storage capacity of the 
   forming black hole.   
If the information carried by the collapsing source is small, 
$\mu$ can be high.  However, it is always finite since the information carried by the energetic object can never be exactly zero. 

  The  observationally-interesting values 
of $\mu$ are $\mu \lesssim 1$, and smaller the better. 
 For such values 
the swift memory burden response affects the classical evolution substantially. 

We have estimated that for astrophysical black holes
obtained by a conventional collapsing matter, 
$\mu \ll 1$. Therefore, in the mergers of 
astrophysical black holes the memory burden response 
is expected to be significant. 

   For PBH ~\cite{Zeldovich:1967lct, Hawking:1971ei,Carr:1974nx,Chapline:1975ojl,Carr:1975qj}   
   formed as a result of a collapse 
   of the Hubble patch in a radiation dominated 
   epoch, the memory burden load is expected to be close to maximal, regardless the specific production mechanism.  
   For the PBH smaller than the Hubble volume, 
   the information load can be sensitive to the precise formation mechanism and must be estimated on case by case basis.  However,  even for such black holes, the load 
   is typically significant for  endowing the PBH with 
   a sufficiently small $\mu$ for giving a significant swift memory burden response in a merger.  It is thereby expected that 
  generic black holes,  both astrophysical or PBH, 
  in mergers must  exhibit the swift memory burden response which affects their classical dynamics. 
  
  The right place to look for the observational manifestations 
 of the swift memory burden effect  is the gravitational wave signals at wavelengths of order $R$.  
 Basically, our prediction is that the higher order correlators, 
 that are sensitive to the information load of a black hole, affect the classical dynamics of perturbations. 
 
  We made some very preliminary estimates of the 
spectrum of the burdened perturbations for the lowest harmonics and the sensitivity to over-all memory burden parameter $\mu$.  Since $\mu$ is an averaged quantity over all memory modes, the estimates must be further refines by studying the sensitivities  with respect to the burdens carried by different spherical harmonics of the memory modes. 
  Their imprints shall then be translated into the corresponding harmonics of the gravitational radiation. 
 More detailed analysis shall be given in \cite{GiaMichael}.
  
   The significance of the swift memory burden effect is that 
 it is predicted to take place in astrophysical 
 black holes of ordinary Einsteinian gravity, without 
 assumption of any new physics.
 It directly follows from the well-accepted 
 premise that black holes are  the most compact (and therefore most efficient) storers of information. 
  Under this premise, the memory burden effect, and in particular its swift manifestation, appears to be inevitable. 
   We see no conceivable way for two pairs 
   of merging black holes, with identical initial characteristics but different information loads, to merge similarly. 

The only miraculous option would be that, due to some new principle, the  evolution of a black hole proceeds 
exactly on null memory-burden surface. 

We are not aware of any known criterion demanding such 
evolution.  For example, classical black holes obey 
all sorts of no-hair theorems \cite{NH1,NH2,NH3, NH4, NH5, NH6, NH7}. However, such theorems 
do not forbid a black hole to carry temporary features,  
i.e.,  the features that after some time fade away. 
 Nor they forbid the quantum features. 
 Therefore the swift memory burden effect is in no conflict 
 with the classical black hole no-hear properties: first,  it is quantum in origin and, secondly, it is temporary. 
   
   For the completeness of the picture and the appreciation 
  of  the universality of the phenomenon,  
    we have demonstrated the presence of the memory burden response 
 in solitons.   We used  as a prototype the  monopole  of high efficiency of information storage constructed in \cite{Dvali:2019jjw}.  Upon the merger of such solitons, the 
 swift memory burden gets activated and influences the classical dynamics.   The effect was shown to be operative for memory modes of both fermionic and bosonic types. 
 
  We have also put forward an outline of an experimental proposal for the laboratory tests of the memory burden
  effect in systems with cold bosons.
   In fact, the original Hamiltonians exhibiting the memory burden effect  \cite{Dvali:2018xpy, Dvali:2020wft} 
   admit straightforward  interpretation in terms of cold bosons.  However, we have illustrated the proposal 
  on a simpler system of the attractive bosons on a ring  \cite{ring, Kanamoto:2008zz}, which was used in  \cite{Dvali:2012en, Dvali:2013vxa, Dvali:2015ywa, Dvali:2015wca, Dvali:2016zqx} 
  for  modelling the critical graviton condensate of  black hole master modes.  Our analysis shows that 
 the memory burden effect and in particular its swift 
 manifestation is potentially testable in table-top labs.  \\

 {\bf Acknowledgements} \\
 
  We thanks Lasha Berezhiani for discussions.

 This work was supported in part by the Humboldt Foundation under the Humboldt Professorship Award, by the European Research Council Gravities Horizon Grant AO number: 850 173-6, by the Deutsche Forschungsgemeinschaft (DFG, German Research Foundation) under Germany’s Excellence Strategy - EXC-2111 - 390814868, Germany’s Excellence Strategy under Excellence Cluster Origins 
EXC 2094 – 390783311.   \\

\noindent {\bf Disclaimer:} Funded by the European Union. Views and opinions expressed are however those of the authors only and do not necessarily reflect those of the European Union or European Research Council. Neither the European Union nor the granting authority can be held responsible for them.\\


\begin{thebibliography}{10}
\bibitem{MichaelZ} 
Series of numerical simulations of mergers of the saturated soliton bubbles have been performed by M.~Zantedeschi.
\end{thebibliography}
\end{document}